\documentclass[final,3p,times,twocolumn,authoryear]{elsarticle}
\usepackage{amssymb}
\usepackage{_macros}
\usepackage{multirow}
\usepackage{graphicx}
\usepackage{booktabs}
\usepackage{adjustbox}
\hypersetup{
    colorlinks,
    linkcolor=black,
    citecolor=black,
    urlcolor=black
}

\newcommand{\name}{AniBalloons}

\newcommand{\rv}[1]{\textcolor{black}{#1}}

\newcommand{\rvn}[1]{\textcolor{black}{#1}}

\journal{International Journal of Human-Computer Studies}

\begin{document}

\begin{frontmatter}



\title{AniBalloons: Animated Chat Balloons as Affective Augmentation for Social Messaging and Chatbot Interaction}

\author[label1]{Pengcheng An}
\affiliation[label1]{organization={School of Design, Southern University of Science and Technology},%Department and Organization
            %addressline={Southern University of Science and Technology}, 
            city={Shenzhen},
            %postcode={518055}, 
            %state={Guangdong},
            country={China}
            }

\author[label1]{Chaoyu Zhang}

\author[label2]{Haichen Gao}
\affiliation[label2]{organization={School of Creative Media, City University of Hong Kong},%Department and Organization
            %addressline={Southern University of Science and Technology}, 
            city={Hong Kong},
            %postcode={518055}, 
            %state={Guangdong},
            country={China}
            }

\author[label3]{Ziqi Zhou}
\affiliation[label3]{organization={School of Computer Science, University of Waterloo},%Department and Organization
            %addressline={Southern University of Science and Technology}, 
            city={Waterloo},
            %postcode={518055}, 
            %state={Guangdong},
            country={Canada}
            }

 \author[label1]{Yage Xiao}

\author[label3]{Jian Zhao\corref{cor1}}

%\author{Anonymous Authors}

\fntext[f1]
{\textbf{\small{See a video demonstration via: \href{https://youtu.be/qYASU2a0ktU}{ClickHere}}}}
% \cortext[cor1]{Anonymized for review}
\cortext[cor1]{Corresponding Author}

\begin{abstract}
Despite being prominent and ubiquitous, message-based communication is limited in nonverbally conveying emotions. Besides emoticons or stickers, messaging users continue seeking richer options for affective communication. Recent research explored using chat-balloons' shape and color to communicate emotional states. However, little work explored whether and how chat-balloon animations could be designed to convey emotions. We present the design of AniBalloons, 30 chat-balloon animations conveying Joy, Anger, Sadness, Surprise, Fear, and Calmness. Using AniBalloons as a research means, we conducted three studies to assess the animations' affect recognizability and emotional properties ($N=40$), and probe how animated chat-balloons would influence communication experience in typical scenarios including instant messaging ($N=72$) and chatbot service ($N=70$). Our exploration contributes a set of chat-balloon animations to complement nonverbal affective communication for a range of text-message interfaces, and empirical insights into how animated chat-balloons might mediate particular conversation experiences (e.g., perceived interpersonal closeness, or chatbot personality).
\end{abstract}


% \begin{highlights}
% \item A longitudinal iterative design process with five professional art therapists. 
% \item A novel AI-infused digital art-making system \name{}.
% % \item results of evaluation in various setups understand AI as a material in art therapy
% \item \rev{Results of an evaluation that assists the understanding of AI as a material in art therapy.}
% \end{highlights}


\end{frontmatter}

%\linenumbers

% \section{}
% \label{}
%\newpage
\section{Introduction}\label{sec:introduction}

Messages are one of the most prominent means of communication in today's society. It is estimated that Short Message Services (SMS) are being used by five billion people around the world\footnote{https://www.vodafone.com/business/news-and-insights/blog/gigabit-thinking/the-evolution-of-the-text-message}. 
Message-based interaction has been evolving and adapting to multiple mobile and wearable devices, such as mobile phones, smartwatches, smart glasses, etc.
The user population of mobile and wearable messaging apps had surpassed three billion\footnote{https://www.businessofapps.com/data/messaging-app-market}, with the majority from WhatsApp, Facebook Messenger, and WeChat, which also allow for cross-device communications. 
%On WhatsApp alone, over 100 billion messages are sent each day\footnote{https://backlinko.com/whatsapp-users}. 
The growing usage of messages has been further propelled after the COVID period \citep{textIncreaseDuringCovid}, revealing its importance in various use contexts, such as connecting distant friends and family \citep{textBenefitInPandemic}, facilitating patient-doctor communication \citep{texthealthcare}, supporting telecommuting \citep{notUpset}, and enabling chatbot-supported businesses\footnote{https://www.uschamber.com/co/good-company/launch-pad/text-based-commerce-surges-during-pandemic}.% and client support\footnote{https://www.drift.com/books-reports/conversational-marketing-trends/\#2021+Report+Key+Findings}.

Despite its ubiquity and prominence, text chats are by nature, rather limited in nonverbal aspects of communication, especially when it comes to nonverbally conveying affective information \citep{emoBalloon,vibemoji}. In face-to-face communication, people could discern rich and nuanced feelings from each other based on numerous nonverbal social signals, including facial expressions, vocal features, bodily motions, etc. \citep{vinciarelli2009socialSignal}. 
Most of these nonverbal aspects are sacrificed in message-based communication \citep{textEmotion}, resulting in less sense of connectedness and presence \citep{stayConnected}, and increased likelihood of miscommunication of emotional states \citep{MiscommunicationEmotion,emoBalloon}. 
\rvn{As suggested by Social Information Processing Theory originally developed by Salancik and Pfeffer (\citeyear{salancik1978social}) and later revisited by Walther  (\citeyear{waltherSIP92}),
Computer-Mediated Communication (CMC) may require more time to foster social relationships compared to face-to-face communication due to the limited channels and absence of non-verbal cues \citep{SIPTheory}; therefore, the inclusion of non-verbal elements may enhance the richness and expressiveness of communication, benefiting the development of interpersonal relationships.}

\rvn{Similarly, according to Social Presence Theory \citep{short1976social}, social presense, or the ``sense of being with another'', is crucial for affording warm and sociable interpersonal interactions. 
Different communication media offer varying levels of social presence, corresponding to their richness of social cues. 
For instance, face-to-face communication is considered the richest medium, followed by video conferencing, phone calls, and then text-based communication, which is less rich due to fewer social cues \citep{yen2008online}.}
As a compensation in practice, message senders commonly use pictographic elements such as emojis/emoticons, in parallel with messages, as an added nonverbal channel to express emotions \citep{emoticonCommunication}. While emoticons predominantly represent facial expressions, there remain a lot more design opportunities for visually communicating emotions by integrating other forms of social signals \citep{vinciarelli2009socialSignal}. For instance, HCI research has explored depicting affects expressed in vocal features using typefaces \citep{emotype} or animated texts \citep{animatedtexts}. Prior research also suggested the promise of incorporating vibrations \citep{vibemoji} or bio-signals \citep{heartRateMessageLiu}. %Additionally, Liu et al. depicted bio-signals to enhance the affective connection between messaging senders \citep{heartRateMessageLiu}.

Beyond augmenting typeface or emoticons, recently, a few studies started exploring \textit{chat balloons} as a novel medium for visually conveying emotions \citep{emoBalloon,bubbleColoring}. Namely, Aoki et al. (\citeyear{emoBalloon}) presented EmoBalloon which visualizes the emotional arousal of a message through the ``explosion'' shape of its encasing chat balloon generated by an ACGAN model. Chen et al. (\citeyear{bubbleColoring}) explored utilizing colors of chat balloons to indicate affects of voice messages (e.g., excitement, anger, sadness and serenity). These two studies suggested a great potential for further probing and designing the affordance of chat balloons in affective communication. 

However, to date, little has been known about whether and how chat balloon \textit{animations} could be designed to convey intended emotions. This under-addressed opportunity has become the primary motivation of our design-driven research. In other design areas, animation has been long utilized for conveying affective cues \citep{lasseter1987principles,de2017taxonomy,chevalier2016animations}: for instance, enhancing audience's affective resonance to data storytelling designs \citep{kineticharts}. Prior work in multimodal emoticons found that simple animations could already be interpreted by users as dynamic affective cues such as body language, and such dynamic affective cues could add a new level of nuance in messages in addition to static affective cues \citep{vibemoji}. It is thereby suggested that exploring the animations of chat balloons may result in a unique and complementary affective channel for messages, and more design research is hence needed.

In this research, following a structured affective animation design process (introduced by \cite{kineticharts}), and a set of design requirements specifically formulated for chat balloons, we have designed \name{}. \name{} is a set of chat balloon animations that could communicate six types of emotions: Joy, Anger, Sadness, Surprise, Fear and Calmness. 
The first five are from the so-called basic emotions that commonly manifested and universally understood \citep{ekman1992argument}. Calmness is for users to express the state of being free from strong emotions. 
we analyzed 230 affective animation examples to extract design patterns for each emotion category, and iteratively designed five animations for each category.

We utilized \name{} as a means of inquiry to answer the following two research questions:
\begin{itemize}
    \item \rv{\textbf{RQ1: \textit{How could chat balloon animations be designed to convey emotions?}}}
    
    \item \rv{\textbf{RQ2: \textit{How would people perceive the emotional properties of the chat balloon animations?}}}
    
    \item \textbf{RQ3: \textit{How might animated chat balloons influence user experience in message-based interaction?}}
\end{itemize}

\rv{A formative, design-driven research (\textbf{Study 1}) was conducted with design professionals to address \textbf{RQ1}.} 
To answer \rv{\textbf{RQ2}, \textbf{Study 2}} assessed the affect recognizability of the designed animations, and their perceived emotional properties according to the valence-arousal emotion model ($N=40$). To answer \rv{\textbf{RQ3}}, we created high-fidelity design mockups to explore two typical yet distinct scenarios of message-based interaction: mobile messaging in \rv{\textbf{Study 3-a}} ($N=72$), and chatbot interaction in \rv{\textbf{Study 3-b}} ($N=70$). In both studies, we took a between-group setup to compare the relevant aspects of participants' experience in a pre-scripted conversation under two conditions: one with animated chat balloons implemented for the message interface, and the other with only static chat balloons. 

Our studies results showed that 80\% of the designed animations were effective in communicating the intended emotions, and the animation designs together covered a variety of valence-arousal parameters, which suggested the great potential of chat-balloon animations in serving as a unique affective channel for messages. Moreover, we found that animated chat balloons could enhance perceived emotional communication quality, nonverbal information conveyance, and the sense of closeness in the context of mobile messaging. And they could also increase the chatbot's likeability, make it funner to interact with, and affect its perceived personality trait. These findings indicated that animated chat balloons could be leveraged to mediate particular aspects of conversation experience in message-based interaction, which opens up a broad range of opportunities for future design and research. 

Our contribution in this paper is thereby two-fold: (1) a set of chat balloon animations that could support nonverbal affective communication for a range of message-based interfaces; (2) empirical insights into how animated chat balloons would influence user experience in instant messaging and chatbot interaction.

\section{Related Work}

\subsection{Affective Enhancement for Messages}
 %exclamation mark, emojis and emoticon and memes etc. research and innovation on emojis and emoticons, and research on how to enhance text message emotioanl communication of chatbots (interpersonal and human-agent)
Due to the inherent limitation of text messages in nonverbally conveying emotions, HCI systems have implemented different approaches to enhance the affective channels in text message-based communication.
Emoticons, or emojis, have been the focus of the majority of research in this area. Emoticons have a long history of augmenting messages with affective information \citep{ScottFahlman,plato}. 
Studies have also set out to understand the emotional states of users based on their usage of emoticons in various contexts such as education \citep{Zhang2017_affectivestates}, software development \citep{Chen2021_emotion}, and public discussions \citep{Hagen2019_emojiuse}.
Studies have also revealed user-led creative use of emoticons emerging in real-world scenarios, such as repurposing emojis \citep{Wiseman2018_repurposingemoji,kelly2015characterising} and replacing text with emojis \citep{zhou2017_wechat}. 
Alongside HCI research, the industry has been continuously introducing new designs of emoticons \citep{newemoji2021,facebookmessenger,Telegram} and new user customization capabilities \citep{applememoji,googlegboard}.

While emoticons have been the primary means of enhancing affective expression in text-based communication, other techniques have also been explored by HCI research.
Studies have examined the use of images such as static or animated memes as a medium for emotional communication.
For example, Kim et al. (\citeyear{messageImage}) proposed a system that recommends images to match the context of a message. Jiang et al. (\citeyear{jiang2017_GIF}) found that animated GIFs can trigger nuanced interpretations and enrich the nonverbal communication experience. DearBoard by Griggio et al. (\citeyear{Griggio2021}) supported the co-customization of image stickers across messaging apps for nonverbal expressions between intimate partners. Beyond images, the visual features of texts, such as typefaces \citep{emotype}, or text animations, have also been explored \citep{emotionalSubtitles}. Emotype \citep{emotype} presented emotional typefaces that mean to match a selected emoticon, to enhance the message's emotionality. Animated texts were also explored to convey nonverbal cues to television viewers with hearing impairment \citep{emotionalSubtitles}. A recent work by Hautasaari et al. leveraged emotional text captions to enhance attendees' communication experience of online conferencing \cite{emoScribe}. Wakey-Wakey \citep{WakeyWakey} proposed an automatic framework that could transfer existing GIFs' aniamtion schemes to text. Liu et al. depicted heart rate data to enhance affective communication in social messaging \citep{heartRateMessageLiu}, and they also developed Animo \citep{AnimoLiu}, a smartwatch-based animated agent that enables users' affective connection. Similarly, Wang et al. (\citeyear{EmotionChatPhysiological}) developed a generative technique to translate bio-signals into text animations. Buschek et al. (\citeyear{ChatAugmentation}) utilized physiological data and contextual factors to customize fonts to augment chats. 

In summary, various methods, such as emoticons, images, and visual elements of text, have been examined to enhance emotional communication for messages. While these elements can all be considered as part of the message's ``content'', less exploration has been done on the universal ``container'' of messages, which is the chat balloons, leaving ample opportunities for design and research.

\subsection{Chat Balloon Related Affective Design}
Chat balloons or text bubbles are common in text-based communication interfaces. Before digital media, speech balloons had played a crucial role in comics and manga, serving not only as carriers for speech, but also as emotional indicators \citep{emoBalloon}. 
Studies by Yamanishi et al. on manga have shown the connection between the type of speech balloon used and the linguistic properties of the speech, as well as the relationship between the shape of the speech balloon and the emotional arousal level of the message \citep{balloonShape}.

Built further upon, Aoki et al. (\citeyear{emoBalloon}) presented the EmoBalloon system, which used explosion-shaped chat balloons to communicate arousal level of the message sender. These explosion shapes were generated by an Auxiliary Classifier GAN (ACGAN). The evaluation results of EmoBalloon suggested that the shape of the chat balloons could accurately convey emotional arousal in text messaging. Besides emotional states, speech balloon shapes were also used to indicate other nonverbal information \citep{peng2018speechbubbles,comicChat}. Chen et al. (\citeyear{bubbleColoring}) explored using colors of voice message bubbles to represent senders' affects such as excitement, anger, or serenity:e.g., a red bubble to indicate an angry message, and an orange bubble to indicate a happy message. 

Besides shape and color, the animation of chat balloons also seems to be a potential affective channel for text-based communication. However, there is limited research on how to design chat balloon animations to convey emotions. The only related exploration was from Shi et al. (\citeyear{voiceAgent}), which used facial expressions, text box movements, and voice waveforms to convey different emotional states of a voice assistant (idle, bored, curious, etc.). However, it is unclear if the text box animations alone could convey the intended emotions or if they can be used across various message contexts. Our work aimed to address this gap by formally assessing the designed 30 affective animations in terms of affect recognizability and emotional properties and exploring their use in two distinct contexts. Overall, while prior studies show the potential of chat balloons to complement nonverbal affective cues for messages, the use of chat balloon animations as an affective medium remains under-explored and requires more design and research.

\subsection{Affective Animation Design for HCI Systems}
The use of animation to convey emotional aspects has a long history in animation theory \citep{thomas1995illusion,louToLife,lasseter1987principles}. In the field of User Interface design \citep{chevalier2016animations} and Data Visualization \citep{de2017taxonomy}, animations are also used as a means of depicting and evoking emotional feelings. The additional temporal dimension of animations allows for the depiction of more nuanced or richer affective states that cannot be conveyed through static media. For example, Sonderegger et al. presented AniSAM and AniAvatar \citep{Sonderegger2016_AniSAMAniAvatar}, a set of animated feedback tools for measuring affective states. The study found that animated forms led to higher accuracy than static forms in measuring participants' arousal, highlighting the value of animations as nonverbal representations of emotions. Animo by Liu et al. (\citeyear{AnimoLiu}) demonstrated how animations of simple geometries could depict the physiological data of smartwatch users and enhance their affective bonding. Another smartwatch application, Significant Otter \citep{Otter} used diverse animations of two otters, influenced by shared biosignals, to express intimacy and affect between partners, despite its primary focus being on biosignal sharing rather than animation design.

HCI research has been continually exploring the design of animations or kinetic motions for GUI components. One of the most renowned works is Kineticons by Harrison et al. (\citeyear{Harrison2011}). Kineticons proposed a large set of animation designs, or a ``kineticon vocabulary'', that could be applied to a wide range of GUI elements. Although Kineticons were not initially designed or assessed for affective communication, there are several motion designs that could be associated with emotional expression (e.g., ``jump wave'', ``jump reach'', ``heart beat'', or ``shake no''). In the study of VibEmoji \citep{vibemoji}, a subset of Kineticons was used to create an animation library for multi-modal emoticons to enrich emotional communication in mobile messaging.
Our research, inspired by Kineticons, aims to provide a collection of chat balloon animations that can support affective expression in a wide range of message-based interaction scenarios. Similarly, in the field of data visualization, Lan et al. recently presented Kineticharts \citep{kineticharts}, which designed a collection of animation effects for bar charts, pie charts, and line charts to enhance the affective expressiveness of data stories. Despite its focus on charts, Kineticharts demonstrates a structured way of designing affective animations that could potentially be applied to other design areas, such as chat balloons.

Research and design related to affective communication often rely on two main types of emotion theories: discrete emotion models and dimensional emotion models. Dimensional emotion models view emotional states as a continuous distribution over an emotional space, whereas discrete emotion models categorize emotions into distinct, generalized categories. For example, Russel and Barrett's valence-arousal model \citep{russell1999core} describes each emotional state in terms of its valence (positive or negative) and arousal (arousing or calm), whereas Ekman's work (\citeyear{ekman1992argument,ekman1992there}) identifies basic emotions that are commonly experienced and universally understood across cultures. While previous affective designs have been assessed based on either discrete \citep{kineticharts,bubbleColoring, AffectiveGarmentFoo}, or dimensional \citep{emoBalloon,vibemoji} views of emotion, in this research, we evaluated our designed affective animations by assessing both affect category recognizability (discrete) and valence-arousal properties (dimensional) to have a comprehensive understanding of the affective affordance of the animated chat balloons.
\section{\rv{Study 1: Creating \name{} through an Iterative Design-driven Research}}

In this section, \rv{adopting a Research-through-Design (RtD) approach \citep{Zimmerman_RtD}. we present our process and outcomes of addressing \textbf{RQ1}}, including the formulation of design requirements, a structured, iterative design process, and finally, the 30 affective animations of \name{} \citep{Aniballoon}.   

\subsection{Design Requirements}
While affective animation design has been widely practiced in HCI sub-domains such as GUI \citep{chevalier2016animations} or Data Visualization \citep{de2017taxonomy,kineticharts}, few studies have explored its practical implication for chat balloons. We thereby formulated the following design requirements to explicate the specificity of our design target and guide our design practice:

\textbf{D1: \textit{The designed chat balloon animations could be applied on various message-based interfaces.}}
This is to make sure the designed animation schemes could be easily \rv{implemented in practice, for instance, when integrated into an existing keyboard interface for users to select and combine with their message on the spot---as similar to Vibemoji \citep{vibemoji} or DearBoard \citep{Griggio2021}. Hence, to make the animations compatible with various message interfaces and scenarios,} the design should be based on a generic chat balloon shape (e.g., rounded rectangles), be compatible with different base colors, and be tested on different types of interfaces (e.g., mobile messaging and chatbot interaction). 

\textbf{D2: \textit{The chat balloon animations are able to be used together with emojis in a complementary way.}}
Emojis and emoticons have been commonly added to messages to convey emotional cues through facial expressions. Our design aims to expand the emotional bandwidth for messages without replacing existing cues. Therefore, it is important that the created chat balloon animations and emojis can work together, completing instead of competing with each other. To achieve this, our design focuses on the motion patterns of chat balloons and the dynamic effects of their abstract decoration, while avoiding any concrete facial features on the balloons, to ensure that the animations will enhance rather than simply repeat the emotional cues provided by emojis when used together.

\textbf{D3: \textit{The designed chat balloon animations should not demand prolonged engagement from users.}} Our goal in designing chat balloon animations is to enhance the emotional expressiveness of messages without demanding users' prolonged engagement. To achieve this, we have kept the repetition cycle of each animation to under four seconds and used simple, abstract shapes in the animation sequences. This allows for a quick understanding of the animation's pattern, which allows for short engagement. Additionally, we have avoided the use of detailed graphic elements and confined each animation to its surrounding area to avoid being intrusive during a conversation. Additionally, we confined each animation to its surrounding area and avoided having ``whole-screen'' animations (e.g., screen effects in Apple iMessage) which might be distracting in a conversation.

\textbf{D4: \textit{The chat balloon animations should not hinder users from reading the message.}} Chat balloon animations meant to convey emotional cues should not sacrifice the readability of the message. This consideration has guided many of our design choices. For example, in animations that feature fast motion, we have chosen to keep the message content static or moving at a slower pace and smaller range (as a design metaphor, we see the chat balloons as containers holding viscous liquid with the message content floating within). Additionally, in animations that include perspective distortion or transition effects, we have extended the interval duration between animation cycles to maintain appropriate readability.  

\subsection{Design Process}
Our design process of \name{} consists of five stages by following the design requirements:

\textbf{Identifying Initial Categories:}
To guide our design, we conducted a design inspiration analysis to identify common patterns used by animation designers to convey emotions. We used a structured workflow introduced in Kineticharts \citep{kineticharts}, which is meant to elicit affective animation patterns from design inspirations. The first step was to choose which emotions to design for, and we narrowed it down to the six "basic emotions" identified by Eckman in the 1970s: \emph{joy, anger, fear, surprise, sadness, and disgust} \citep{ekman1992there}. These emotions are universally recognized and commonly experienced in everyday life, regardless of cultural or language differences \citep{ekman1992there,ekman1992argument}. These basic emotions are also widely referenced by HCI research, e.g., \citep{kineticharts,AffectiveGarmentFoo}. Additionally, we included an extra category of emotional state, calmness, to give users the option to express feelings of being free from strong emotions.

\textbf{Collecting Initial Inspirations:}
With the initial emotion categories formed, we began collecting design inspirations. We used a process similar to that outlined in Kineticharts \citep{kineticharts} and first took an inclusive approach to build an archive of existing animation design examples. We searched popular design platforms such as Dribbble, Behance, and Pinterest, where animation designers post and exchange their motion graphic designs. Keywords (synonyms) related to the seven emotion categories were used. \rv{Animations deemed capable of effectively communicating the designated emotions were included. This process involved a review by two researchers; an animation was included if either researcher found it adequate. In the subsequent \textit{Refining Inspirations} phase, a more selective approach was applied to further refine and filter the samples.} Regarding the emotion of disgust, we found a significant lack of design references. 
Given our adopted affective design approach \citep{kineticharts} primarily relies on extracting design patterns from rich examples,
we chose to prioritize the six categories that seemed best suited for this approach, while leaving out disgust for future development. In total, we collected 336 motion graphic designs in the GIF or MOV format.

\begin{figure}[!t]
  \centering
  \includegraphics[width=\linewidth]{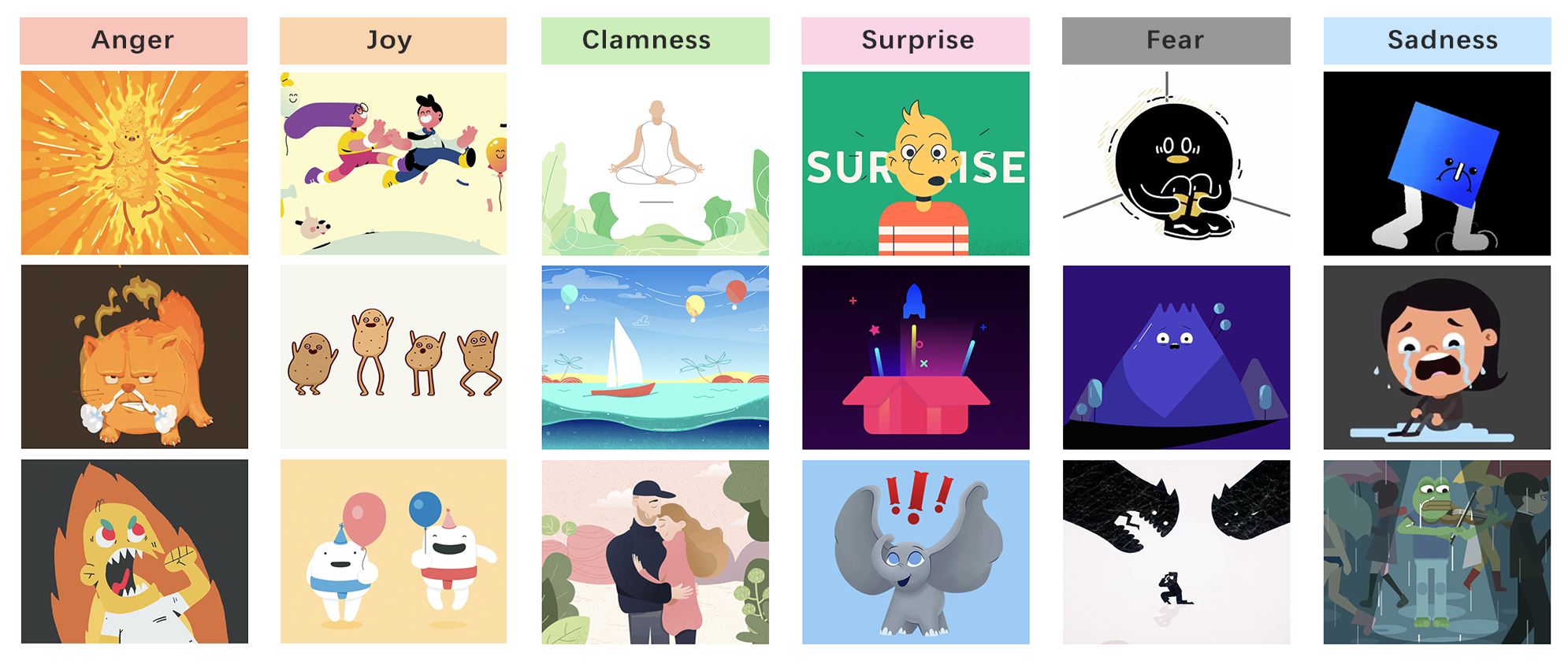}
  % \vspace{-7mm}
  \caption{Samples of collected design inspirations. }
  \label{fig:inspiration}
\end{figure}

\textbf{Refining Inspirations:} 
An exclusive approach was then taken to refine the design inspirations for the finalized emotion categories: \textbf{joy}, \textbf{anger}, \textbf{fear}, \textbf{sadness}, \textbf{surprise}, and \textbf{calmness}. Two designers from our research team carried out this refinement for each category. Our design emphasized the role of motion and nonverbal dynamic effects in conveying emotions, as a complement to facial expression-based emojis (\textbf{D2}). Therefore, we excluded examples in which the emotion was mainly communicated by texts or static facial expressions, and the kinetic motions or dynamic effects contributed little. Some typical examples included a word ``JOY'' with each letter animated but their motion did not evoke a sense of joy; or, a slowly moving angry-face emoji in which anger cannot be discerned from its movement. The two designers examined each category separately and marked designs potentially for exclusion, then discussed and reached final decisions for exclusion. In the end, 230 design examples remained after the refinement, including 52 inspirations for joy, 50 for anger, 39 for fear, 39 for sadness, 50 for surprise, and 38 for calmness. Examples of these design inspirations can be seen in \autoref{fig:inspiration}.

\textbf{Analyzing Inspirations:}
An open coding method \citep{moghaddam2006coding} was adopted to extract affective animation design patterns from our collected inspirations. Following prior studies \citep{motionDesignSpace,animatedDataGraphics} while guided by our design requirements, we coded the inspirations based on three aspects: (i) the main object's motion; (ii) the decorative dynamic effects; and (iii) timing. We first annotated around 40\% of the inspirations, from which the initial codes emerged. Then, each category was coded by two designers separately, and new codes could still be added during the process. The two designers then worked together to finalize the coding results and cluster the codes into design patterns in these three aspects. (i-iii). For example, in terms of the main object's motion (i), we found that ``trembling'', ``huddling'' and ``jumping'' were often used to animate a character's fear. Decorative elements' dynamic effects (ii), such as shaking dashed lines (or ``yure-sen'' in Japanese) or fluctuating silhouettes were often used to convey fear. Timing (iii) was especially helpful in distinguishing subtle differences in motion designs. For example, ``jumping'' was a common motion for expressing both joy and fear but joyous jumping often repeated multiple times and had longer "slow-in", while fearful jumping often happened only once and rather suddenly. These extracted patterns helped us to effectively characterize how the emotions can be depicted by design elements.

\textbf{Iterative Design and Expert Involvement:} 
Using the extracted design patterns, three designers from the research team designed chat balloon animations for the six emotion categories iteratively. They first carried out divergent design experimentation on whether or how the animation patterns (e.g., main character/object's kinetic motion and decorative dynamic effects) could be translated into motions of a "generic" chat balloon in a rounded rectangular shape (\textbf{D1}). They found that sometimes a single pattern was enough to convey the desired emotion, while other times a combination of patterns was needed. The outcomes included designs inspired by both single patterns and pattern combinations, deliberately balancing expressiveness with unobtrusiveness, to avoid demanding prolonged engagement (\textbf{D3}).

After generating a variety of design ideas in the divergent stage, a convergent approach was taken. The design team selected a set of promising design samples that effectively conveyed the intended emotions, while also representing a diversity of design patterns. In this stage, the team paid more attention to the patterns of timing to fine-tune the temporal features of each design sample and optimize its emotional communicability.

To further develop the designs, we carried out three expert review sessions with three professional animation designers. We chose two from technology companies in North America and one running her own animation design agency in Asia. All experts were women and had more than five years of industry experience. The sessions were conducted via video call, where we introduced the background of our project and presented the design samples category by category. The experts were encouraged to freely provide feedback at any time, either after seeing a single sample or the whole category. All experts were interested in the concept of using chat balloon animations to convey emotions and offered valuable feedback. 

The experts' suggestions for improving the designs covered a wide range of aspects, such as the metaphor of motion (e.g., animating the body language of ``arms'' and ``legs'' via the corners of the balloon), specific choices for the decorative dynamic effects (e.g., conveying anger via spitting ``fire'' instead of ``steam''), and the fine-tuning of temporal patterns (e.g., adjusting the speed of ``shaking'' to more vividly express fear). Moreover, the experts also helped with optimizing the readability of the textual message for each animation design, discarding a few design ideas that employed graphical effects in the background of the textual message (\textbf{D4}). As a result, six to seven selected design samples in each category were further detailed and polished. The team then conducted another round of divergence-convergence design iteration, creating variations of the designs based on the expert's suggestions, comparing and selecting the most effective designs for further fine-tuning. Finally, five representative animations for each emotion category were carefully polished to create the final design collection.

\begin{figure*}[!t]
  \centering
  \includegraphics[width=\linewidth]{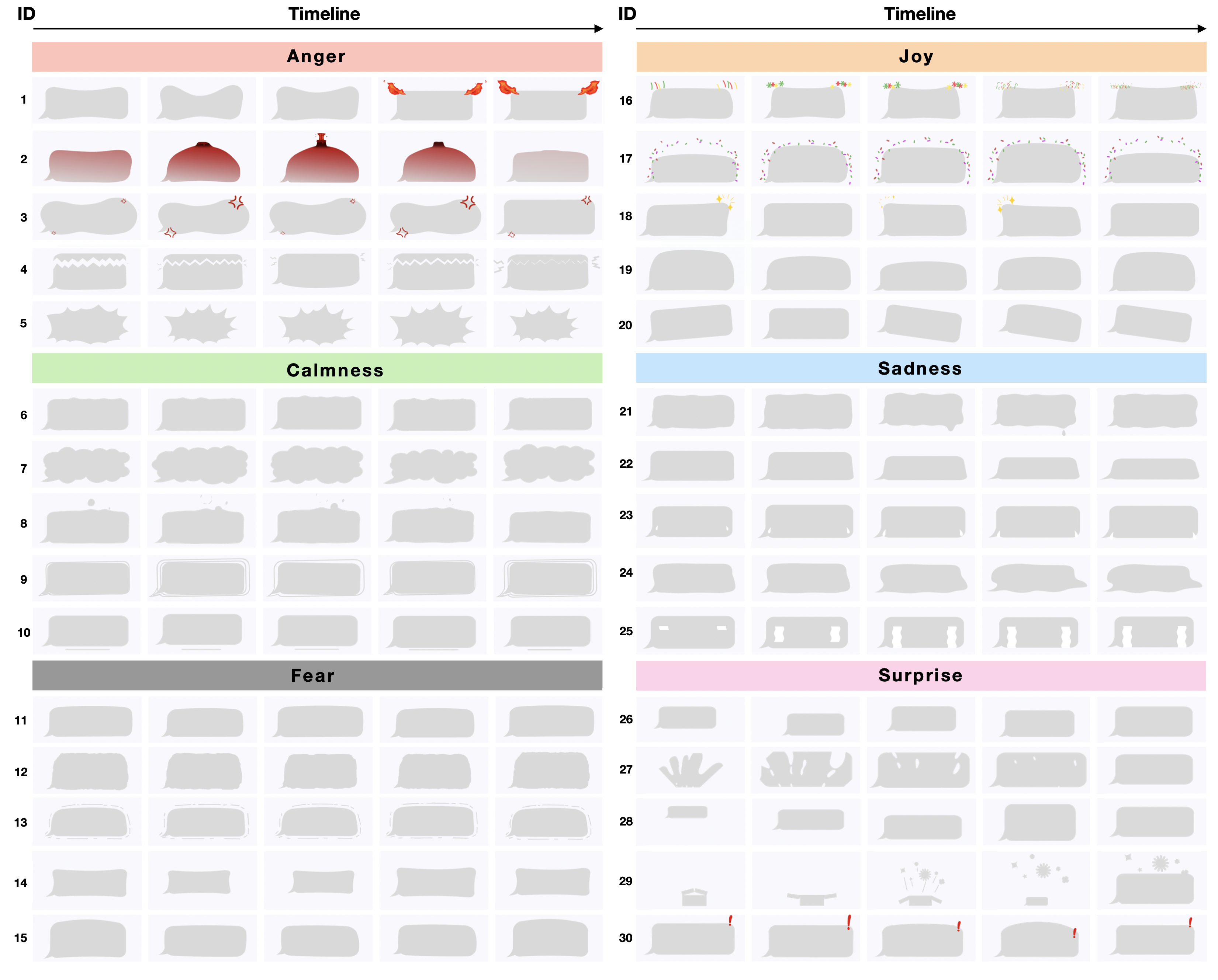}
  % \captionsetup{justification=centering}
  % \vspace{-8mm}
  \caption[caption]{
  The 30 designed \name{}:\\
  {\footnotesize{\tabular[t]{rlllll}
  \textbf{Anger:} & 1-Fire-spitting & 2-Erupting & 3-Clenching & 4-Shouting & 5-Exploding\\
  \textbf{Calmness:} & 6-Rippling & 7-Breathing & 8-Bubbling & 9-Drifting & 10-Floating\\
  \textbf{Fear:} & 11-Quivering & 12-Shuddering & 13-Shaking & 14-Huddling & 15-Recoiling\\
  \textbf{Joy:} & 16-Celebrating & 17-Cheering & 18-Shining & 19-Hailing & 20-Swinging\\
  \textbf{Sadness:} & 21-Tear-shedding & 22-Lying-down& 23-Weeping& 24-Melting &25-Whining\\
  \textbf{Surprise:} & 26-Zooming & 27-Splashing & 28-Popping & 29-Unboxing & 30-Springing
  \endtabular}}
  
  \textbf{\small{See an anonymized video demonstration via: \href{https://youtu.be/qYASU2a0ktU}{ClickHere}}}
  }
  %\caption[caption]{This is the caption\\\hspace{\textwidth}This is the second line}
  \label{fig:AniBalloons}
\end{figure*}

\subsection{\name{}}
As illustrated in \autoref{fig:AniBalloons} (also see the demo video \href{https://youtu.be/qYASU2a0ktU}{ClickHere}), the final designs of \name{} consist of 30 unique animations. It is important to note that the names given to the designs are solely for referencing purposes in this paper and do not limit the potential use contexts of each animation. The following is a brief overview of the \name{} designs for each emotion category:

\textbf{Anger.} The five animations in this category are \textit{Fire-spitting}, \textit{Erupting}, \textit{Clenching}, \textit{Shouting}, and \textit{Exploding}. We found that a common animation pattern for expressing anger is squeezing a certain part of the body with tension (e.g. shrinking shoulders, tucking chin, or balling hands into fists). Both \textit{Fire-spitting} and \textit{Clenching} use this pattern and include dynamic effects such as spitting fire or displaying "cross-popping veins." \textit{Erupting} uses the motion of expanding the body (e.g., puffing out chest), along with a volcano eruption effect. \textit{Shouting} features the motion of angry shouting and mumbling, paired with the effect of shark teeth biting. In \textit{Exploding}, the chat balloon expands and then "explodes."
  
\textbf{Calmness.} This category includes \textit{Rippling}, \textit{Breathing}, \textit{Bubbling}, \textit{Drifting}, and \textit{Floating}. The motion patterns for calmness often involve metaphors related to water, air, and the state of being floating or weightless. For example, \textit{Rippling} was inspired by the motion of gentle waves. Similarly, \textit{Drifting} depicts the motion of floating on water. \textit{Bubbling} gives the impression of bubbles in soda or other carbonated drinks. \textit{Breathing} combines the metaphor of a cloud with the motion of slow deep breathing. \textit{Floating} illustrates the motion of (a meditating character) floating in the air and moving slightly and slowly up and down.
  
\textbf{Fear.} This category includes \textit{Quivering}, \textit{Shuddering}, \textit{Shaking}, \textit{Huddling}, and \textit{Recoiling}. Different types of ``trembling'' motions were predominantly seen in the inspirations. However, through detailed analysis, subtle variations were identified and incorporated into the designs. For example, \textit{Quivering} is solely based on a trembling motion, while \textit{Shuddering} and \textit{Shaking} also integrate dynamic effects such as "fluctuating silhouette" and "shaking dashed lines" respectively. \textit{Huddling} combines trembling with body shrinking, a common pattern in fear-evoking animations. Lastly, \textit{Recoiling} combines a sudden jump with trembling, similar to the "jump scare" moment commonly seen in inspirations.
  
\textbf{Joy.} The five animations are \textit{Celebrating}, \textit{Cheering}, \textit{Shining}, \textit{Hailing}, and \textit{Swinging}.
\textit{Celebrating} features the motion of stretching both arms upward and incorporates a firework effect. \textit{Cheering} and \textit{Hailing} both employ the typical joyous motion of jumping repetitively, with the confetti effect added in \textit{Cheering} and a different motion pace in \textit{Hailing}. \textit{Shining} uses the motion of stretching one side of the body and arm alternately, accompanied by a star-shining effect. Lastly, \textit{Swinging} shows the motion of jumping and swinging left and right, a pattern commonly seen in inspiration examples.

\textbf{Sadness.} The five designs in this category are \textit{Tear-shedding}, \textit{Lying-down}, \textit{Weeping}, \textit{Melting}, and \textit{Whining}. Our analysis revealed that motions related to crying are often used to convey sadness in animations. The design of \textit{Tear-shedding} captures the abstract impression of tears welling up in the eyes. \textit{Weeping} combines the motion of sobbing with the dynamic effect of tears drops. \textit{Whining} illustrates the impression of breaking down in tears with an abstract tear-streaming-down effect. In contrast, \textit{Lying-down} represents the motion of characters laying back or lying down in sadness. Lastly, \textit{Melting} captures the unique motion of anthropomorphic objects sadly collapsing or melting.
  
\textbf{Surprise.} The designs for surprise are \textit{Zooming}, \textit{Splashing}, \textit{Popping}, \textit{Unboxing}, and \textit{Springing}. The patterns extracted for surprise often involved sudden appearance or proximity. \textit{Zooming} is designed to convey the feeling of something quickly approaching, represented by the balloon rapidly expanding. \textit{Splashing} evokes the sensation of liquid being splashed on a surface. \textit{Popping} features the motion of jumping out and bouncing on the ground. \textit{Unboxing} uses the metaphor of a message coming out of a gift box. Lastly, \textit{Springing} combines an exclamation mark effect with the motion of jumping in surprise.

\section{\rv{Study 2}: Emotion Recognizability and Emotional Properties of \name{}}

Utilizing \name{} as a research means, we aim to address the question of whether or how the designed chat balloon animations could convey emotions, and how people would perceive their emotional properties (\textbf{RQ2}). To do so, we conducted a study with 40 participants to evaluate the 30 designed animations with two-fold objectives: 

\textbf{Objective 1: \textit{Assessing the emotion recognizability of the designed motion effects of \name{}.}}
This involved assessing whether or to what extent participants could identify the intended emotion conveyed by each chat balloon animation without any hint from the textual content of a message. As per prior research, an affective design is considered effective when recognition accuracy exceeds 50\% \citep{kineticharts,Ma2012GuidelinesFD} (i.e., most respondents could recognize the intended emotion without any hint from the texts).

\textbf{Objective 2: \textit{Gathering the perceived emotional properties of the designed motion effects of \name{}.}}
This objective aimed to understand how the designs of \name{} are perceived in terms of valence and arousal. By collecting data on the perceived emotional properties of each design, we can identify the range of emotions that \name{} can effectively convey and how they can complement other emotional cues (e.g., emoticons) and identify potential areas for future design developments.

\subsection{Methods}

\begin{figure*}[tb]
  \centering
  \includegraphics[width=\linewidth]{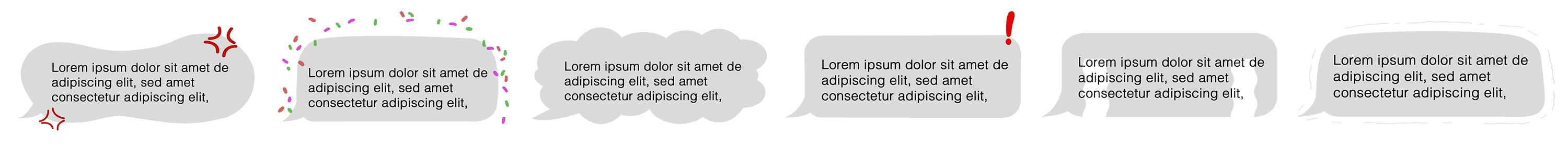}
  % \vspace{-8mm}
  \caption{\name{} used for evaluation: rendered on chat balloons with grey base color and pseudo Latin texts}
  \label{fig:greyballoons}
\end{figure*}

\textbf{Stimuli:}
The 30 designed motion effects of \name{} have been used as stimuli. To ensure that the evaluation was not influenced by the base color of the chat balloons, which can vary in actual usage cases (\textbf{D1}), each design was rendered on a grey chat balloon. However, the color of the dynamic decorative effects (e.g., see the confetti or question mark in \autoref{fig:greyballoons}) were not changed, as these would remain in the same color in actual usage cases. 
Additionally, to prevent the content of the message from influencing participants' perception, while still providing a sense of a complete chat balloon, we used Lorem ipsum, a type of pseudo-Latin text, as the placeholder message in the balloon.

\textbf{Procedure:}
The 40 participants were recruited by the research team via word of mouth, aged from 18-44 (95\% aged from 18-34), with 42.5\% women and 57.5\% men. A web-based survey interface was created for the evaluation. After a participant completed consent and demographic information, the survey interface demonstrated each of the 30 designs one by one to the participant (rendered on an identical grey balloon with the same placeholder text), and meanwhile, ask the participant to fill in a few evaluation questions. The designs were shown to each participant in a randomized order to avoid the order effect biasing their evaluation. For each design, a participant was first asked to choose which emotion the sender of the rendered message seemed to convey, with seven options provided: Joy, Surprise, Sadness, Anger, Calmness, Fear, and Other. When Other was opted for, as an optional request, they could input a better-suited answer in a text field. Secondly, the participant was asked to rate their perception of valence (the extent of being pleasant-unpleasant) and arousal (the extent of being calm-exciting) expressed by the rendered message in two seven-point scales. Afterwards, 8 participants were randomly invited for a 15-min follow-up remote interview to talk about their responses to the survey.
\rv{The interviews were transcribed verbatim and subjected to a thematic analysis \citep{thematicAnalysis} where two researchers engaged in collaborative inductive coding. Initially, they annotated transcripts to highlight pertinent quotes, essential concepts, and initial patterns. Through recurrent discussions, they formulated a coding scheme, under which quotes were organized and grouped into emerging themes, forming a hierarchical structure. These themes were reviewed and finalized collaboratively with the research team.}

\begin{figure*}[!tb]
  \centering
  \includegraphics[width=\linewidth]{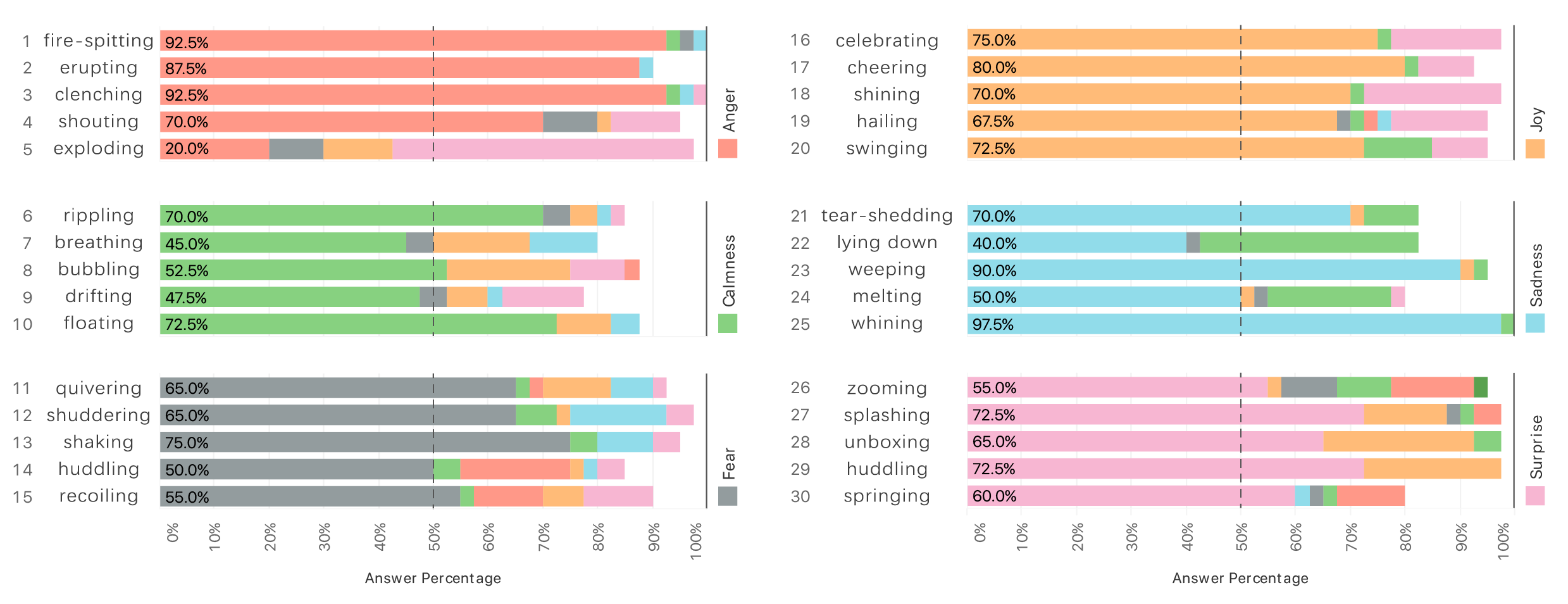}
  % \vspace{-8mm}
  \caption{Emotion recognizability of \name{}}

  \label{fig:recognizability}
\end{figure*}

\subsection{Emotion Recognizability of \name{}}

%- a large image about recognizability test results

As \autoref{fig:recognizability} shows, 24 out of 30 animations (80\%) have achieved a recognition accuracy of greater than 50\%, which are considered effective in affective design according to prior works \citep{kineticharts,Ma2012GuidelinesFD}. Among them, 20 designs reached an accuracy of 65\% or higher.
In the categories of Joy and Surprise, all the designed affective animations could be successfully recognized by the majority of the participants. And in the category of Fear or Anger, four out of five animations could be correctly recognized by the majority. And in the category of Sadness or Calmness, three out of five animations could be correctly recognized by the majority.

\rv{We further conducted a subsequent analysis through a binomial test, which showed that for 93\% of the designs (28 out of 30), the intended emotion was identified significantly more often than would be expected by chance alone (p $<$ .001). Moreover, for these 93\% designs, the selection rates for alternative emotions did not significantly differ from chance. This underscores a consensus among participants regarding the conveyed emotion.}

Six animations' accuracy did not exceed 50\%, including \textit{Exploding} (20\%) from Anger, \textit{Breathing} (45\%) and \textit{Drifting} (47.5\%) from Calmness, \textit{Huddling} (50\%) from Fear, \textit{Lying-down} (40\%), and \textit{Melting} (50\%) from Sadness. However, when looking at the responses from participants who chose "Other," some interpretations were closely related to the intended emotion. For example, for \textit{Drifting} (Calm), 7.5\% of participants interpreted it as meditation, no emotion, or sleepiness. 5\% of participants thought \textit{Huddling} (Fear) conveyed anxiety. And 7.5\% of participants saw \textit{Melting} (Sadness) as expressing disappointment, frustration, upset, and feeling down. If these related interpretations are considered correct answers, 27 (90\%) animations could be considered effective in conveying the intended emotion. \rv{The participants' interpretations also included alternative meanings to the conveyed emotions.
For instance, \textit{Exploding} was identified by a participant to convey "importance", and \textit{Breathing} was identified by two participants as "thinking". These alternative interpretations could serve as inputs for refining or expanding the designs of \name{}.
}

\textbf{Qualitative Results:} 
While most of the designs' affective expressions were commonly agreed upon by the participants, there were a few animations that elicited diverged interpretations. For instance, \textit{Exploding} (designed for Anger), was interpreted as Surprise by 50\% of participants, and \textit{Lying-down} (designed for Sadness) was perceived as Calmness by 40\% of participants. 
Several participants provided explained how they would apply \textit{Exploding} to a message to express surprise: e.g., ``\textit{like `I will be home tomorrow!'}'' Another participant stated that she would use it to express ``emphasis'', such as ``\textit{[reminding people:] don't forget to do it}''. ``Lying-down'' was associated with tiredness by multiple participants: e.g., ``\textit{I'm so tired! So many things to do next week.}'' Interestingly, several participants stated that they would use \textit{Bubbling} and \textit{Drifting} to express relaxation or leisure. Notably, the designs with lower agreement were not viewed as ``bad''. Instead, the participants gave their individual interpretations and envisaged when and how they would like to use them to express feelings beyond the original design intent (echoing findings in \citep{Wiseman2018_repurposingemoji,kelly2015characterising,vibemoji}).

\begin{figure*}[]
  \centering
  \includegraphics[width=0.67\linewidth]{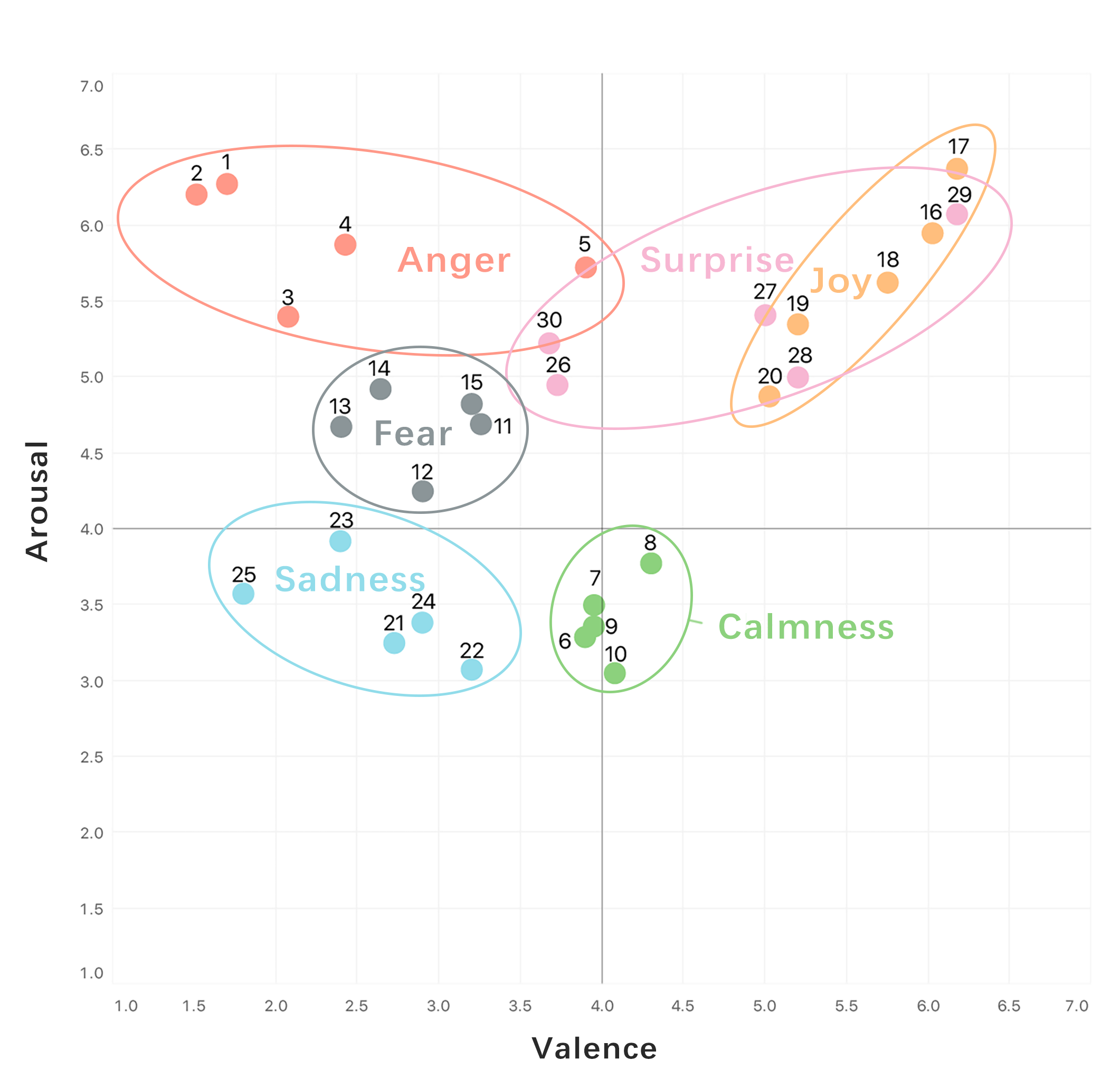}
  \includegraphics[width=0.67\linewidth]{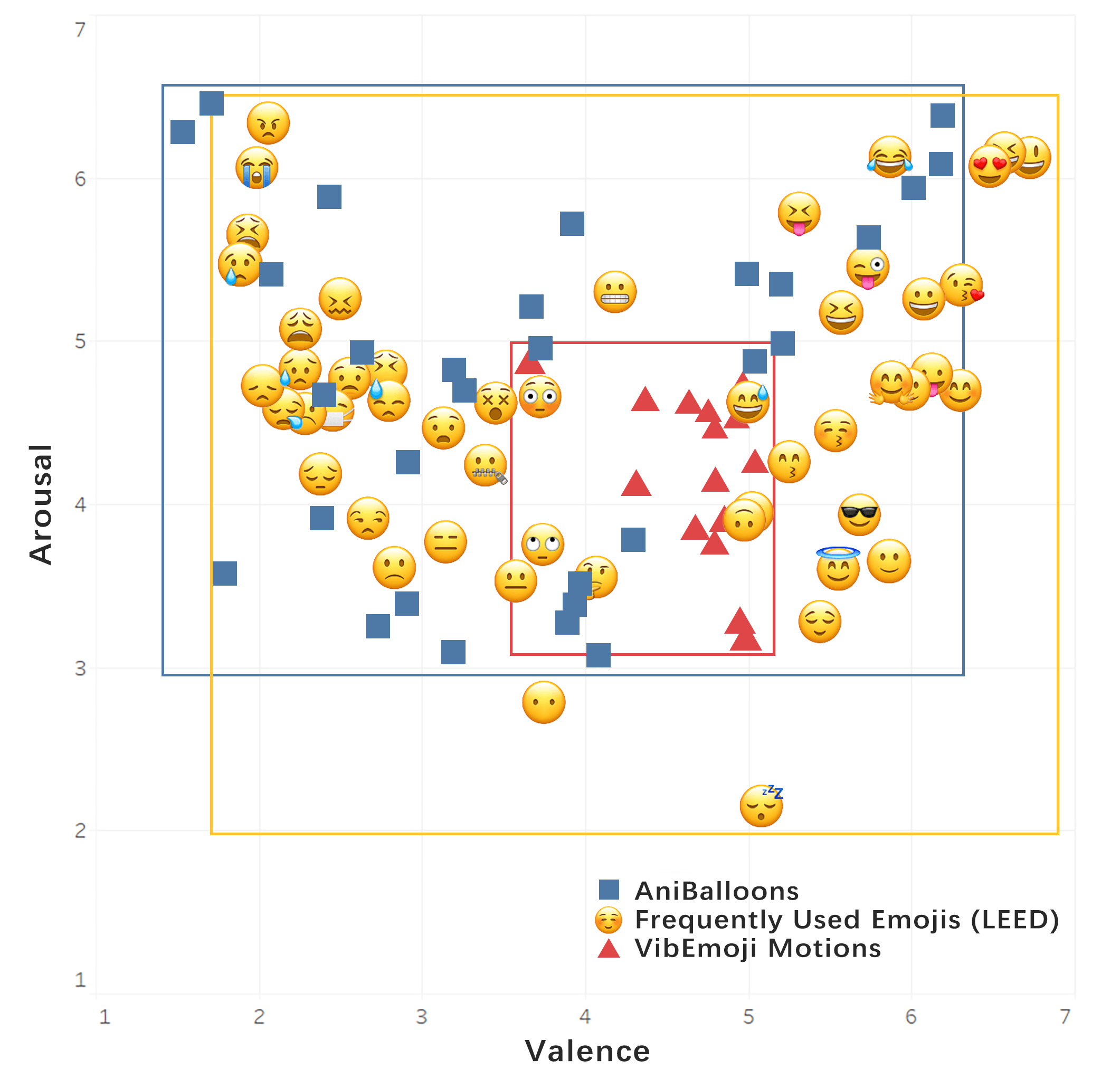}
  % \vspace{-3mm}
  \caption{Emotional properties of \name{}. TOP: \name{}' distribution on valence-arousal plane. BOTTOM: \name{}'s distribution in reference with frequently used emojis and VibEmoji \citep{vibemoji} motions.}

  \label{fig:Properties}
\end{figure*}

\subsection{Perceived Emotional Properties of \name{}}
This subsection unveils the nuanced emotional properties of the designed animations and how they relate to one another on the valence-arousal plane.

The distribution of the designs on this plane is illustrated in \autoref{fig:Properties}. 
Each emotion category occupies a specific range on the plane (marked by the ellipses on \autoref{fig:Properties}), providing users with nuanced options for expressing a certain emotion. 
The Joy designs are broadly distributed in the positive-arousing section of the plane, with the most positive and arousing design being \textit{Cheering} and the least being \textit{Swinging}.
The Anger and Fear designs are located in the negative-arousing section, with \textit{Erupting} and \textit{Shaking} perceived as expressing the strongest negative feelings from the two categories.
The Sadness designs are in the negative-calm section, with \textit{Whining} perceived as the most negative.
The Surprise designs are primarily located in the arousing section and span across the neutral and positive regions on the valence axis (as Surprise can evoke a wide range of valence), with \textit{Unboxing} conveying the highest positivity and \textit{Zooming} and \textit{Springing} conveying valence-neutral surprise. 
The Calmness designs are located in the calm section of the plane and around the neutral region on the valence axis, which aligns with the intended design of conveying a feeling of being free from strong emotions. 
The distribution of the designs also aligns with the results of the emotion recognizability evaluation, as some participants identified Joy in some Surprise designs and vice versa. This may be due to the overlap between the two clusters in real-life, as expressing happy surprise often involves both the emotions of surprise and joy. Similarly, The designs that evoked different interpretations (such as \textit{Exploding} and \textit{Lying-down}) were located near the border regions of other clusters. By utilizing such proximity-based associations evident in the distribution data, animations can be recommended based on their nuanced emotional properties, rather than solely relying on predefined emotion labels.

To further understand the affective affordance of \name{}, in \autoref{fig:Properties}, we visualize the distribution of \name{} in reference to frequently used emojis \citep{rodrigues2018lisbon}, as well as a prior set of animations used for multimodal emoticons \citep{vibemoji}. 

In the figure, 50 frequently used (facial) emojis from the Apple IOS platform were plotted based on a survey by the Unicode Consortium \footnote{https://home.unicode.org/emoji/emoji-frequency/}, with their valence-arousal properties taken from the Lisbon Emoji and Emoticon Dataset, i.e., LEED \citep{rodrigues2018lisbon}. Additionally, a prior set of animations from the VibEmoji system, consisting of 15 motion designs selected from Kineticons and with valence-arousal data gathered in \citep{vibemoji}, were plotted in the figure.

\rv{To be noted, the LEED dataset serves merely as a contextual backdrop in \autoref{fig:Properties} to illustrate the distribution of commonly used emojis within the valence-arousal space. Our goal was to furnish the audience with a referenced overview to better understand the positioning of Aniballoons in the emotional landscape.} 
As can be seen in \autoref{fig:Properties}, the distribution of the 30 \name{} designs covers a relatively wide range of emotional parameters, which is on par with the range covered by frequently used emojis and include the range covered by the VibEmoji animations. To be noted, the VibEmoji animations were designed to enhance multimodal emoticons, rather than communicating emotions alone, thus we use their data as a reference point rather than making a direct comparison. The 50 frequently used emojis, on the other hand, cover a range of emotions commonly used in everyday communication, and the large overlap between the two suggests that \name{} can also support a wide range of affective expression in daily contexts. However, as can be seen in the figure, there may be a need for more designs in the positive-calm phase of the emotional space in the future to support affective expression in this area.

\textbf{Qualitative Results:} 
The interviews with participants provided additional insights into how \name{} can complement existing methods of emotional expression, such as emojis and stickers. 
As echoing our design purpose, a few participants mentioned that \name{} could complement emojis because emojis were static while \name{} convey feelings through motions and dynamic effects. A participant pinpointed that the meaning of emojis tends to be context-dependent, whereas ``\textit{the chat balloon animations are more straightforward, thus less likely to be misinterpreted}''. Additionally, participants pointed out that \name{} can be integrated seamlessly into any message that has a chat balloon, whereas ``\textit{stickers require an extra message. So sometimes I hesitate whether to send a sticker [...] but this [chat balloon] is part of your message.}'' Another interesting difference mentioned was ``\textit{the chat balloon animations are more universal, their usage are not restricted to gender, personality, age [...]}'' Whereas when using emojis or stickers, people tend to select the appropriate ones (e.g., to match their skin color, gender, or culture). This is likely due to that emojis and stickers are designed with specific facial features or figures, while \name{} were designed based on extracted motion patterns and relatively abstract effects.
\section{\rv{Study 3-a}: Potential Effects of Animated Chat Balloons on Messaging Experience}
This study explores the potential use of animated chat balloons in mobile instant messaging, a common context for text-based communication. The research question is how animated chat balloons would impact the conversation experience in this context (\textbf{RQ3}). A between-group study was thereby conducted with 72 participants to examine the influence on participants' experience when applying animated chat balloons in an instant messaging scenario.

\subsection{Methods}

\textbf{Hypotheses:} 
\rv{Based on our research questions, four hypotheses were formulated. Among them, H1 has been based on our results of Study 2, and H2 and H4 were inspired by prior related work. We now address these hypotheses and rationales behind each:}

\textbf{H1: \textit{Applying animated chat balloons can influence perceived affective communication quality.}} 
Findings from Study 2 suggest that \name{} could effectively convey emotional feelings. Therefore, we hypothesize that applying these designs in a mobile messaging scenario can influence the perceived quality of affective communication, such as increasing the perceived emotional expressiveness and liveliness of the conversation.
  
\textbf{H2: \textit{Applying animated chat balloons can influence perceived conveyance of nonverbal information.} }
Nonverbal interactions play an important role in enhancing the sense of social presence in computer-mediated communication \citep{de2001socialpresence,yen2008online}, and have been greatly sacrificed in messaging due to the lack of nonverbal cues \citep{emoBalloon}. Since \name{} are based largely on embodied motions, we hypothesize that they enhance the conveyance of nonverbal information and make the conversation feel more like face-to-face interaction.
  
\textbf{H3: \textit{Applying animated chat balloons would not influence perceived message comprehensibility.} }
Animated chat balloons are designed to nonverbally convey emotional feelings, while the comprehensibility of messages depends more on verbal features. Therefore, we expect that the use of \name{} will not have a significant impact on the perceived comprehensibility of messages.
  
\textbf{H4: \textit{Applying animated chat balloons could influence perceived closeness to the conversation partner.}}
Sharing emotions can strengthen the relationship and sense of connectedness between people \citep{RIME2020127sharingEmotion}. We will explore whether the use of animated chat balloons in messaging can mediate participants' perception of interpersonal closeness to the conversation partner.

\begin{figure*}[tb]
  \centering
  \includegraphics[width=\linewidth]{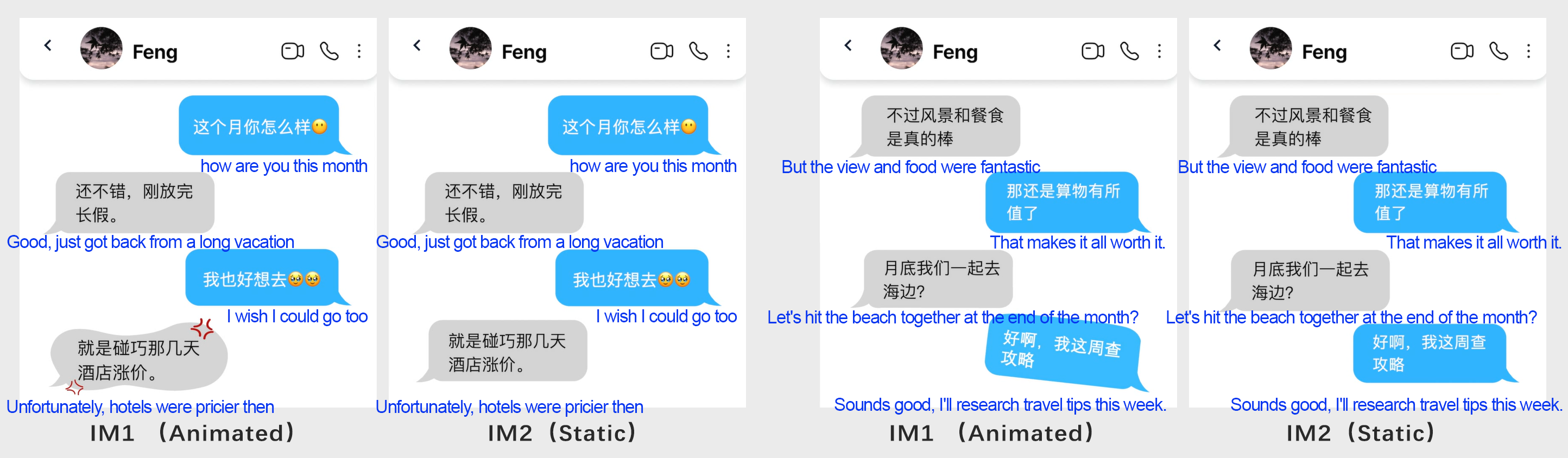}
  % \vspace{-7mm}
  \caption{Screenshots of the mobile messaging mockup interfaces (with the English translation added).}

  \label{fig:IM}
\end{figure*}

\textbf{Stimuli:}
In order to evaluate the potential impact of using \name{} in mobile instant messaging, two versions of high-fidelity mockups, called \textbf{IM1} and \textbf{IM2}, were created (see \autoref{fig:IM} and the supplemental video). Both mockups apply the same pre-scripted conversation and visual design, with the only difference being that certain messages in \textbf{IM1} are augmented with \name{}, while all messages in \textbf{IM2} are displayed in static text balloons.

The conversation script in the mockups was designed to resemble a typical mobile messaging exchange between a user and a conversation partner discussing the partner's recent holiday trip. The conversation begins with a greeting and inquires about the partner's recent life, then the partner shares that they just returned from a holiday trip and proceeds to describe the details of the trip. Throughout the conversation, the partner mainly expresses their feelings about the trip, while the user's perspective mostly gives comments. The conversation ends with the partner and the user agreeing to travel together in the future. The conversation was divided into thirteen messages, with seven from the user and six from the partner.

In \textbf{IM1}, the chat balloons of nine messages (five from the partner and four from the user) were augmented with \name{} animations to convey emotions related to joy, surprise, anger, calm, sadness, and fear. These emotions were chosen based on the context of the conversation, for example, joy is conveyed when the partner expresses pleasure about the food and scenery at the destination, and when the user agrees to travel with the partner. Surprise is expressed when the user hears about the unexpected increase of the hotel price, and when the partner suggests traveling together. Anger is shown when the partner talks about the hotel price. Calm is rendered when the user admires the partner's destination. Sadness is communicated when the partner mentions the destination being very crowded. Fear is conveyed by the user when reacting to the crowdedness of the place. In \textbf{IM1}, only one \name{} animation is displayed at a time to avoid visual clutter and to help users focus on the most recent emotion being conveyed.

\begin{table}[!t]
  \caption{Measurements of \rv{Study 3-a}.}
  % \vspace{-3mm}
  \centering
  \label{tab:S2}
  \small
  \begin{tabular}{llp{0.7\linewidth}}
    \toprule
    
    % \midrule
    \textbf{H1} & \textbf{IM-RS1} & Our conversation was emotionally rich.\\
    & \textbf{IM-RS2} & Our conversation was lively.\\
    & \textbf{IM-RS3} & Our conversation was expressive.\\
    & \textbf{IM-RS4} & My partners’ emotions were not clear to me.\\
    & \textbf{IM-RS5} & My emotions were not clear to my partner.\\
    \midrule
    \textbf{H2}& \textbf{IM-RS6}  & Our conversation included rich nonverbal information.\\
    & \textbf{IM-RS7} & Some nonverbal information supported our communication.\\
    & \textbf{IM-RS8} & I felt as if I could picture my partner’s expressions and reactions at the other end.\\
    & \textbf{IM-RS9} & It isn’t at all like a face-to-face conversation.\\
    \midrule
    \textbf{H3} & \textbf{IM-RS10} & It was easy to understand
my partner.\\
    & \textbf{IM-RS11} & My partner found it easy to understand me.\\
    \midrule
    \textbf{H4} & \textbf{IOS} & Inclusion of Other in the Self Scale \citep{aron1992inclusion}\\
  \bottomrule
\end{tabular}
\end{table}

\textbf{Measurements:}
To assess the hypotheses, a questionnaire has been constructed. As \autoref{tab:S2} shows, Five items (\textbf{IM-RS1} to \textbf{IM-RS5}) \rvn{have been used to evaluate participants' affective communication experience of the conversation as a whole (\textbf{H1})}, including its emotional richness, liveliness, expressiveness, and \rvn{the perceived mutual understanding of emotions (understandability of Partner's emotions and understandability of one's own emotions to Partner).} \textbf{IM-RS4} and \textbf{IM-RS5} were selected from the ``perceived affective understanding" cluster of Networked Minds Social Presence Measure \citep{networkedMinds}.

Four items (\textbf{IM-RS6} to \textbf{IM-RS9}) were used \rvn{to specifically evaluate the nonverbal information conveyed in the conversation (\textbf{H2})}, including the richness of nonverbal information, perceived usefulness of nonverbal information, the ability to picture reactions from the partner, and the resemblance of face-to-face communication. \textbf{IM-RS8} and \textbf{IM-RS9} were developed based upon two subjective attitude statements from IPO Social Presence Questionnaire \citep{de2001socialpresence}.

Two items (\textbf{IM-RS10} and \textbf{IM-RS11}) were used to assess participants' \rvn{perceived mutual comprehensibility of the conversation content (\textbf{H3})}. These two items were selected from the category ``perceived message understanding'' of the Networked Minds Social Presence Measure \citep{networkedMinds}.

Finally, the Inclusion of Other in the Self (\textbf{IOS}) \citep{aron1992inclusion} Scale has been adopted to assess participants' perceived closeness to the conversation partner (\textbf{H4}). IOS is a standard single-item scale that measures how close a participant feels to another individual or group \citep{aron1992inclusion}.

\rvn{There are subtle yet significant distinctions to be clarified for a few items that seem related. Namely, a message interaction may be expressive (\textbf{IM-RS3}), but this may not automatically enable participants to vividly imagine nonverbal cues as if in direct, face-to-face communication (\textbf{IM-RS8}). Additionally, richer nonverbal cues may not inherently translate to greater perceived closeness (\textbf{IOS}). Similarly, \textbf{IM-RS4} and \textbf{IM-RS5}
are confined to the transmission of emotions, whereas \textbf{IM-RS10} and \textbf{IM-RS11}
encompass the broader comprehension of conversation content.}
All above-mentioned measurements have been constructed as seven-point Likert scales.
\rv{The Mann-Whitney U test, as a non-parametric alternative to the independent t-test, was used to analyze our data due to that it is suitable for determining if two populations differ significantly without assuming a normal distribution for the data and with a relatively limited sample size.}

\textbf{Procedure:}
The study recruited 72 participants, primarily through word of mouth, who were aged 18-59 (with 90\% aged 18-44). The participants were 61.1\% women and 38.9\% men. They were randomly divided into two groups of 36, with age and gender balance, to experience either \textbf{IM1} or \textbf{IM2}. Participants filled out a consent form and demographic information at the beginning of the study. They were then asked to imagine themselves in a scenario where they were chatting with a person from their mobile messaging contacts. They were then provided with the pre-scripted design mockup of either \textbf{IM1} or \textbf{IM2} (both running for 1 min 23 sec) and asked to fill out the questionnaire based on their experience. After the questionnaire data was gathered, 10 participants from the \textbf{IM1} group were randomly selected for a 15 min online interview, in which they were also shown \textbf{IM2} and asked to comment on their experience and questionnaire responses.
\rv{The interviews were transcribed and underwent a thematic analysis following the same steps as described in Study 2.}

\begin{figure}[tb]
  \centering
  \includegraphics[width=\linewidth]{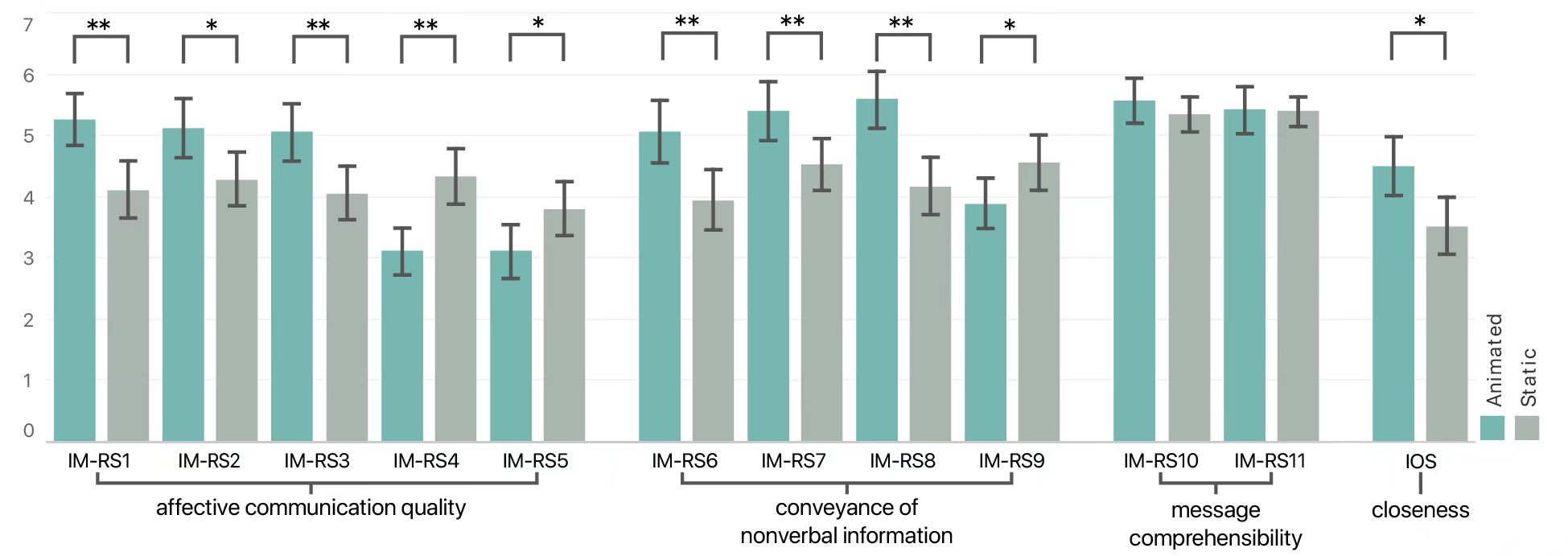}
  % \vspace{-7mm}
  \caption{Results of \rv{Study 3-a}.}

  \label{fig:results2}
\end{figure}

\subsection{Results}
\textbf{Perceived Affective Communication Quality (H1):} In order to assess \textbf{H1}, null hypothesis H0 was tested: the animated chat balloons would not have influence on any of the measurements of perceived affective communication quality. As \autoref{fig:results2} shows, Mann-Whitney U tests suggested that applying animations of \name{} led to significantly better scores compared to the static (no animation) condition in all five measurements (\textbf{IM-RS1} to \textbf{IM-RS5}), including emotional richness (M = 5.25 $>$ 4.11, p = .002), conversation liveliness (M = 5.11 $>$ 4.28, p = .019), reverse-scored emotion understandability of Partner (M = 3.11 $<$ 4.33, p $<$ .001), and reverse-sored emotion understandability of Self to Partner (M = 3.11 $<$ 3.80, p = .026). Therefore, the null hypothesis is rejected, and \textbf{H1} is supported. 

\textbf{Perceived Conveyance of Nonverbal Information (H2):}
To evaluate \textbf{H2}, null hypothesis H0 was tested: the animated chat balloons would not have an effect on the measurements of perceived conveyance of nonverbal information. Mann-Whitney U tests showed that applying animated chat balloons led to significantly better scores compared to the no-animation condition in all four measurements (\textbf{IM-RS6} to \textbf{IM-RS9}), i.e., richness of nonverbal information (M = 5.06 $>$ 3.94, p = .005), perceived usefulness of nonverbal information (M = 5.39 $>$ 4.53, p = .007), ability to picture Partner's reactions (M = 5.58 $>$ 4.17, p $<$ .001), and reverse-scored resemblance of face-to-face conversation (M = 3.89 $<$ 4.56, p $<$ .049). As a result, the null hypothesis is rejected, and \textbf{H2} is supported.

\textbf{Perceived Message Comprehensibility (H3):}
Our hypothesis \textbf{H3} expects no significant effect from chat balloon animations, thereby it aligns with the null hypothesis H0: the animated chat balloons would have no significant influence on the measurements of message comprehensibility. As \autoref{fig:results2} shows, Mann-Whitney U tests suggested that applying animations of \name{} resulted no significant effect on any of the two measurements (\textbf{IM-RS10} and \textbf{IM-RS11}): comprehensibility of Partner (M = 5.56), and comprehensibility of Self to Partner (M = 5.42), in comparison with the no-animation condition (M = 5.33, 5.39, respectively). The null hypothesis cannot be rejected.

\textbf{Perceived Closeness to the Conversation Partner (H4):}
To examining \textbf{H4}, null hypothesis H0 is tested: the animated chat balloons would not influence the perceived closeness to the conversation partner. Mann-Whitney U tests indicated that applying animations of \name{} led to significantly higher perceived closeness measured by the Inclusion of Other in the Self (\textbf{IOS}) Scale (M = 4.50 $>$ 3.53, p = .011). H0 is rejected, and \textbf{H4} is supported.

\textbf{Qualitative Results:}
the comments from the interviewed participants further explained why applying animated chat balloons improved the perceived quality of affective communication. Several participants stressed the importance of expressing emotions during conversations, e.g., ``\textit{When I am chatting with other people, I hope others could feel my emotion [...] so that the conversation doesn't feel too mechanical.}'' And they felt that ``\textit{with the [animated] balloons, you are able to express much more emotion, and I can clearly feel the emotion from the other side.}'' Moreover, they experienced that the emotion conveyed by animated chat balloons were `\textit{`lively}'', ``\textit{intuitive}'' and ``\textit{natural}''. Regarding the animations' effects on the conveyance of nonverbal information, a participant explained that with the animated balloons, ``\textit{it adds to a sense of face-to-face interaction with my friend. In my mind, I could simulate her tone and state when she was saying that, I could imagine her doing exactly the same action [as the animation] when she talks to me.}'' Another participant thought that the animated balloons enabled spontaneous affective comprehension prior to verbal content: ``\textit{[the animations] were intuitive. You don't even need to read the text message to know what the other side wants to express.}'' No issues were reported with message comprehensibility and the participants did not consider the animations hindered the message readability. And they agreed with our design choice that only the latest animation on the interface is rendered to avoid visual clustering. Finally, the feedback from the participants helped us understand how the affective animations led to a stronger sense of closeness to the conversation partner: e.g., ``\textit{when communicating with people close to me, I would try my best to let them feel what I want to express}'', and ``\textit{adding emotional expression will make me feel that he has taken a step further and he values our conversation a lot.}'' For this reason, the majority of the participants stated that they would like to use \name{} with people close to them, such as their family, friends, or significant other.

\section{\rv{Study 3-b}: Potential Effects of Animated Chat Balloons on Chatbot Experience}
Chatbot is another important context that could benefit from affective enhancement of messages (e.g., a client service chatbot uses animated balloons to convey affective feedback to the customer). Therefore, we conducted a study with 70 participants to continue probing \textbf{RQ3} in this context: whether and how applying the animated chat balloons to messages would influence users' conversation experience with a chatbot.

\subsection{Methods}

\textbf{Hypotheses:}
\rv{In light of our research questions, we have formulated four hypotheses among which H1 and H2 were inspired by Study 3-a and H4 was inspired by prior related work. These hypotheses and the rationales behind them are:}

\textbf{H1: \textit{Enabling a chatbot to use animated chat balloons could influence participants' attitude towards the chatbot.}} 
\rv{In this study, our measure of `attitude' specifically refers to the chatbot's fun and likability.} As revealed in Study 3-a, participants considered the messages augmented by \name{} to be fun and engaging, we thereby hypothesize that people might find a chatbot more fun and likable when its text balloons are augmented with animated affective cues than when they are static.  
  
\textbf{H2: \textit{Enabling a chatbot to use animated chat balloons could increase the perceived emotional expressivity of the chatbot.}} 
As reported in Study 3-a, \name{} could enhance the emotional expressivity of a conversation in an instant messaging scenario. We thereby hypothesize that the animated balloons could also enhance a chatbot's emotional expressivity, \rv{which is defined in our measure by the extent to which the chatbot is perceived as human-like in its emotional capacity and its ability to express those emotions.}
  
\textbf{H3: \textit{Enabling a chatbot to use animated chat balloons would not influence the perceived verbal communication ability of the chatbot.}} 
\rvn{Since \name{} has been designed to enrich the emotionality of messages, we hypothesize that \name{} would not have significant impacts on the perceived ability of chatbots to communicate pragmatically.}
  
\textbf{H4: \textit{Enabling a chatbot to use animated chat balloons could influence the perceived personality of the chatbot.}} 
Personality is essential in the emotional design of chatbots and it is often mediated by verbal aspects \citep{kang2018chatbotPurpose,ruane2020personality}. By evaluating whether and how \name{} could influence the perceived chatbot personality, we aimed to probe the potential of using animated chat balloons to mediate chatbot personality. We adopted the widely used personality measure of TIPI based on the Big-Five personality model \citep{TIPI}.

\begin{figure*}[tb]
  \centering
  \includegraphics[width=\linewidth]{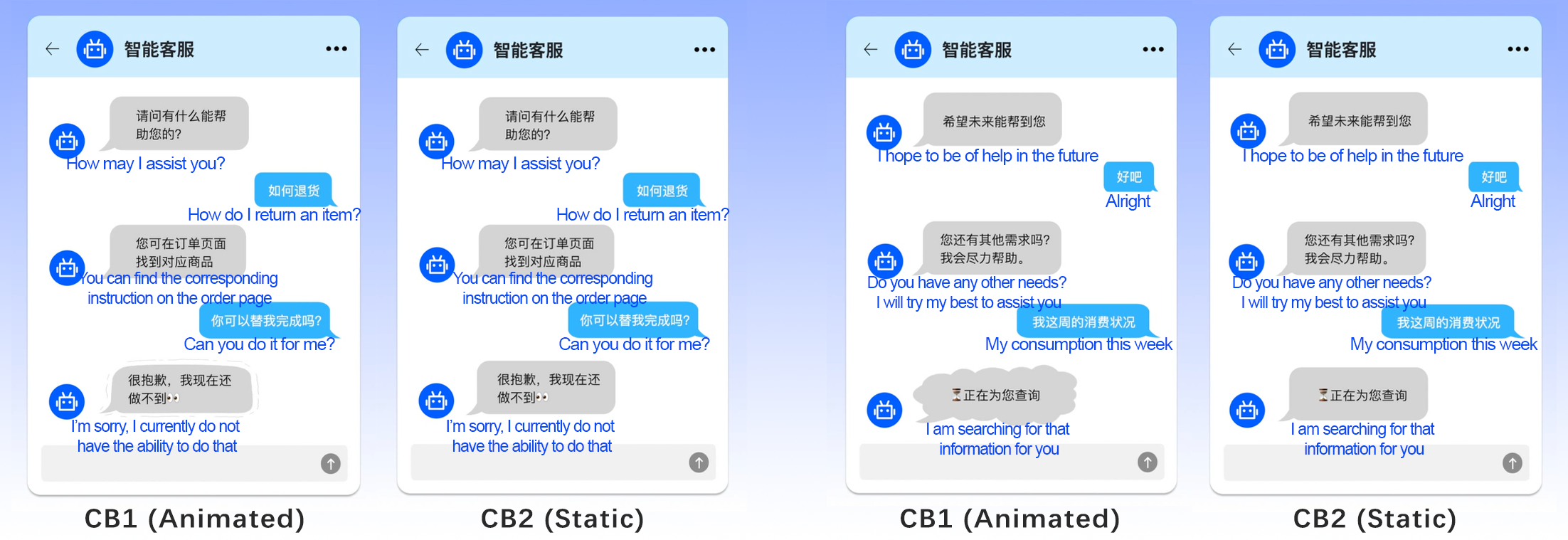}
  % \vspace{-7mm}
  \caption{Screenshots of Chatbot mockup interfaces (with the English translation added).}
  \label{fig:CB}
\end{figure*}

\textbf{Stimuli:}
To evaluate the potential impact of \name{} on the chatbot communication experience, we created two versions of high-fidelity mockups of a customer service chatbot interface: \textbf{CB1} and \textbf{CB2} (see \autoref{fig:CB} and the supplemental video). Both mockups use the same conversation script and visual design. The only difference is that the chatbot in \textbf{CB1} is able to use chat balloons augmented with \name{} to provide affective feedback to the user, while the chatbot in \textbf{CB2} only uses static chat balloons.

The script of the mockups was designed to resemble a typical conversation with a client service chatbot on e-commerce platforms. The script consisted of two common scenarios that would occur in such a conversation. In the first part, the user asks the chatbot about the process of returning a product, and the chatbot explains the steps. Then the user asks the chatbot to complete these steps for them, but the chatbot apologizes and states that it is unable to complete the task but hopes to be able to assist more in the future. In the second part, the chatbot asks if the user needs further assistance and the user asks the chatbot to check their purchase history. The chatbot provides a summary and reminds the user that their expenses have exceeded the monthly budget. Afterwards, the user has no more requests and the chatbot politely concludes the conversation. The conversation was divided into 22 chat balloons, with sixteen from the chatbot and six from the user.

In \textbf{CB1}, based on the context of the conversation script, thirteen chat balloons from the chatbot have been augmented with \name{} to convey joy, calmness, fear, sadness, and surprise. Joy is expressed by the chatbot when it greets the user, fulfills a request, or bids farewell. Calmness is conveyed when the chatbot is retrieving requested information. Fear is expressed when it is unable to fulfill a request or when it sees that the user's expenses have exceeded the budget. Sadness is shown when the chatbot apologizes to the user. Surprise is displayed when the chatbot first appears. Similar to the mockup \textbf{IM1} in Study 3-a, in \textbf{CB1}, only one animation is displayed at a time, with previous animations becoming static to avoid visual clutter and help users focus on the current emotion being conveyed.

\begin{table}[!t]
  \caption{Measurements of \rv{Study 3-b}.}
  \label{tab:S3}
  % \vspace{-3mm}
  \small
  \centering
  \begin{tabular}{llp{0.6\linewidth}}
    \toprule
    %\textbf{H1} & \textbf{ID} & \textbf{Item}\\
    % \midrule
    \textbf{H1} & \textbf{CB-RS1} & The chatbot was fun to interact with.\\
     & \textbf{CB-RS2} & I found the chatbot likeable.\\
    \midrule
    \textbf{H2} & \textbf{CB-RS3} & the chatbot seemed like a human with its own emotions.\\
    & \textbf{CB-RS4} & I felt the chatbot could express its emotion.\\
    \midrule
    \textbf{H3} & \textbf{CB-RS5} & It is easy to tell the chatbot what I would like to do.\\
    & \textbf{CB-RS6} & The chatbot explained gracefully that it could not help me.\\
    \midrule
    \textbf{H4} & \textbf{TIPI1-TIPI10} & Ten-Item Personality Inventory\citep{TIPI}\\
    \bottomrule
\end{tabular}
\end{table}

\textbf{Measurements:}
A questionnaire has been formulated to evaluate our hypotheses. As shown in \autoref{tab:S3}, the participants' general attitude towards a chatbot (\textbf{H1}) was assessed by two items: fun to interact (\textbf{CB-RS1}), and likeability (\textbf{CB-RS2}). The perceived emotional expressivity (\textbf{H2}) of a chatbot was evaluated by two items: exhibition of human-like emotions (\textbf{CB-RS3}) and emotion expression ability (\textbf{CB-RS4}). \rvn{Two items were used to assess whether the perceived ability of a chatbot to communicate pragmatically hinges on the use of emotional balloons
(\textbf{H3}): ability to understand the desires communicated by users (\textbf{CB-RS5}), and ability to gracefully explain its own limitations (\textbf{CB-RS6}).} Four of these items (\textbf{CB-RS1,2,5,6}) were selected from the item pool used by Chatbot Usability Scale \citep{borsci2022CUS}, while two (\textbf{CB-RS3\&4}) were developed by us. To evaluate the perceived personality (\textbf{H4}) of a chatbot, we used the Ten-Item Personality Inventory (\textbf{TIPI} \citep{TIPI}), a standard scale that consists of five pairs of items. Each pair of TIPI measures a personality trait suggested by Big-Five model \citep{TIPI}: Extraversion (\textbf{TIPI1\&6}), Agreeableness (\textbf{TIPI2\&7}), Conscientiousness (\textbf{TIPI3\&8}), Emotional Stability (\textbf{TIPI4\&9}), and Openness to Experiences (\textbf{TIPI5\&10}). All measurements took the form of 7-point Likert scales. \rv{The Mann-Whitney U test was employed due to its suitability for addressing data without necessarily satisfying a normal distribution and with a limited sample size.}

\textbf{Procedure:}
The 70 participants were recruited through word of mouth, aged from 18-44 (90\% aged 18-34), with 58.6\% women, and 41.4\% men. They were randomly divided into two groups of 35 with age and gender balance.
Each participant first completed a consent form and provided demographic information, and then they were provided with the pre-scripted design mockup of either \textbf{CB1} or \textbf{CB2} (both running for 1 min 23 sec). After that, they were asked to fill out the questionnaire based on their experience. Afterwards, 10 participants from the \textbf{CB1} group were selected for a 15-minute online interview, in which they were shown \textbf{CB2} and asked to provide additional feedback. \rv{A thematic analysis was performed to analyze the interview transcript, employing the same method as detailed in Study 2.}

\begin{figure}[tb]
  \centering
  \includegraphics[width=\linewidth]{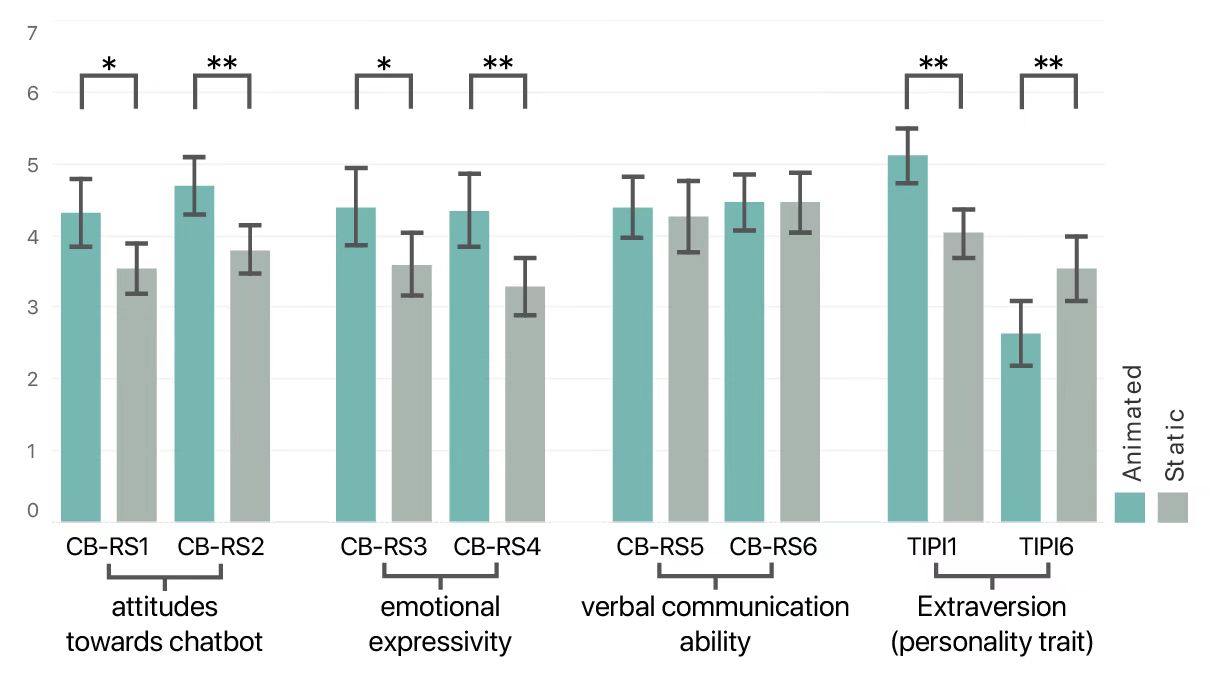}
  % \vspace{-7mm}
  \caption{Results of \rv{Study 3-b}.}
  \label{fig:results3}
\end{figure}

\subsection{Results}

\textbf{General Attitude Towards the Chatbot (H1):}
To evaluate \textbf{H1}, we tested null hypothesis H0: the animated chat balloons would not have any influence on the measurements of participants' general attitude towards the chatbot. As \autoref{fig:results3} illustrates, Mann-Whitney U tests suggested that applying animated chat balloons let to significantly better scores than no-animation condition in the two measurements (\textbf{CB-RS1} and \textbf{CB-RS2}), i.e., fun to interact (m = 4.31 $>$ 3.54, p = .018) and likeability (M = 4.69 $>$ 3.80, p = .002). The null hypothesis is thus rejected, and \textbf{H1} is supported.

\textbf{Perceived Emotional Expressivity (H2):}
To assess \textbf{H2}, null hypothesis H0 was tested: the animated chat balloons would not have any influence on the measurements of perceived emotional expressivity of the chatbot. Mann-Whitney U tests showed that applying animated chat balloons led to significantly better scores than the no-animation condition in both measurements (\textbf{CB-RS3} and \textbf{CB-RS4}): showing human-like emotions (M = 4.40 $>$ 3.60, p = .023), and ability to express emotions (M = 4.34 $>$ 3.29, p = .003). As a result, null hypothesis H0 is rejected, and \textbf{H2} is supported.

\textbf{Perceived Verbal Communication Ability (H3):}
Our hypothesis \textbf{H3} aligns with null hypothesis H0: the animated chat balloons would not have a significant effect on the measurements of the chatbot's verbal communication ability. As \autoref{fig:results3} shows, Mann-Whitney U tests suggested that applying animations of \name{} led to no significant effect on both measurements (\textbf{CB-RS5} and \textbf{CB-RS6}): ease of communication (M = 4.40), and graceful explanation when unable to help (M = 4.45), in comparison with the no-animation condition (M = 4.26, 4.25, respectively). H0 cannot be rejected.

\textbf{Perceived Personality Traits (H4):}
\rv{
To assess \textbf{H4}, the null hypothesis H0 was tested: the animated chat balloons would not influence any of the personality trait measurements: the Ten Item Personality Inventory (TIPI).
TIPI is a brief measure for the Big Five personality traits, comprising two items for each of the five dimensions: Extraversion, Agreeableness, Conscientiousness, Emotional Stability, and Openness.
As \autoref{fig:results3} illustrates, Mann-Whitney U tests suggested that applying animated chat balloons led to significantly higher scores in the personality trait measurements of Extroversion (\textbf{TIPI1} and \textbf{TIPI6}) than no-animation condition, including chatbot being extroverted and enthusiastic (M = 5.11 $>$ 4.03, p $<$ .001), and (reverse-scored) chatbot being reserved and quiet (M = 2.63 $<$ 3.54, p = .007). However, no significant difference was observed in other personality traits (Conscientiousness, Emotional Stability, Openness to Experience) except for Agreeableness, in which only one measurement (TIPI7) showed a significant difference: chatbot being sympathetic and warm (M = 4.54 $>$ 3.38, p = .012). The null hypothesis is hence rejected, and \textbf{H4} is supported.
}
\textbf{Qualitative Results:} 
\rv{
Regarding their general attitude towards the chatbot, a few participants mentioned that they would be more interested to interact with a chatbot applying animated chat balloons than static ones: e.g., ``\textit{It was very fun experience, and I would like to chat with it for a longer time [...]}'' With this increased interst, they might also be more patient with the chatbot: ``\textit{with the chat balloon animations, I might be more patient in waiting for the chatbot's response.}'' Specifically, several participants appreciated the calmness animations used when the chatbot was retrieving requested information, since "\textit{it made the process less boring}'' or ``\textit{it could reduce the anxiety while waiting}''. 
Similar to the instant messaging context, the animated balloons enhanced the emotional expressivity of the chatbot, and according to the participants, this made the chatbot appear more like a human than conventional chatbots: e.g., ``\textit{I felt that the chatbot with animations is more lively, more like a human assistant.}'' ``\textit{because it is in motion, it feels like a real person}'', and ``\textit{I can intuitively feel that the chatbot might be a little emotional [when it failed to help]}'' And such increased emotionality seemed to be a valuable part of user experience: ``\textit{most of the time, chatbot client support cannot really solve my problem. Therefore, if it could bring me at least some emotional feedback, it is already useful.}'' As shown above, the animated chat balloons did not influence the chatbot's perceived verbal ability. However, they seemed to make the chatbot more excusable and empathized when it failed to help users. E.g., ``\textit{It was crying when it cannot help, and I felt sorry for it}'' or ``\textit{It's cute [...] It cannot help me, but is still cute [...] and I really want to excuse it.}'' This suggests a potential opportunity for further research on how nonverbal emotional expression (such as animated message balloons) could mediate users' empathy and tolerance towards conversational agents. Lastly, the participants' comments offered experiential insights echoing our quantitative finding that the animated chat balloons influenced the perceived personality of the chatbot.
Without the animations the chatbot felt like ``\textit{indifferent}'', ``\textit{regular customer support}''. And with the affective animations, people ``\textit{may feel a little bit of temperature}''. And ``\textit{It feels like a relatively young person who does not have a lot of experience but is rather enthusiastic and positive about the work.}'' This implied the need for deeper and more detailed research into how various strategies of implementing affective animations could affect specific personality traits. 
}
\section{Discussion}
% Despite being one of the most ubiquitous and prominent means of communication in today's world, text messages are rather limited in nonverbally conveying affective states \citep{emoBalloon}. 
% A stream of HCI work has thus been focused on enriching the affective expression channels for text message-based communication \citep{emotype,vibemoji,textEmotion,emotionalSubtitles,EmotionChatPhysiological}. 
The prominence and ubiquity of message-based interaction have made chat balloons a universal type of interface component across mobile and wearable platforms.
A few recent studies explored the use of chat balloon shape and color to convey emotions in messages \citep{emoBalloon,bubbleColoring}, but little research has examined the use of chat balloon animations. 
Our design-driven research (Study 1) aims to tackle this under-explored opportunity by creating \name{}, a set of 30 chat balloon animations that convey six emotions: Joy, Anger, Calmness, Fear, Sadness, and Surprise (\textbf{RQ1}). Using \name{} as a vehicle of inquiry, we further address two research questions regarding users' perception (\textbf{RQ2}) and their experiences (\textbf{RQ3}) with chat balloon animations.
%Whether and how chat balloon animations could be designed to convey emotions; and how people would perceive their emotional properties (\textbf{RQ1}). And how animated chat balloons might influence user experience in text message-based communication (\textbf{RQ2})?

In regard to \textbf{RQ2}, our Study 2 showed that even without any context from the message content, 24 designed chat balloon animations were effective in communicating the intended emotions (recognized by the majority of the participants \citep{kineticharts,Ma2012GuidelinesFD}), with 20 designs reached an accuracy of 65\% or higher. Moreover, our analysis of the designs' perceived emotional properties on the valence-arousal plane revealed that each emotion category includes animations that span a certain range, allowing for nuanced differences in affective expressions. And the overall distribution range of \name{} compared to the distribution of commonly used emojis suggests that the designs can largely cover the affective expressions needed in daily communication. 

In regard to \textbf{RQ3}, our Study 3-a and 3-b examined the effects of animated chat balloons on conversation experience in two common contexts of message-based interaction: mobile messaging and chatbot interaction. The results revealed that the use of animated chat balloons in mobile messaging improved the perceived quality of affective communication, specifically by increasing the emotional richness, liveliness, expressiveness, and mutual understanding of emotions in the conversation. Additionally, the animated chat balloons augmented the perceived conveyance of nonverbal information, by making participants feel that they had a better understanding of the partner's reactions and that the conversation felt more like a face-to-face interaction. However, the animated chat balloons did not have a significant impact on the comprehensibility of the message itself. Furthermore, the results demonstrated that the animated chat balloons increased participants' perceived interpersonal closeness to the partner. \rv{These findings, echoing the Social Information Processing Theory \citep{SIPTheory}, suggest that chat balloon animations may offer new opportunities to enhance the nonverbal, affective communication experience (which has been largely restricted in mobile or wearable messaging \citep{textEmotion,heartRateMessageLiu,emoBalloon}) and mediate interpersonal closeness in social applications.}

In Study 3-b, it has been found that the animated chat balloons could significantly influence the participant's attitude to chatbot, making it more enjoyable to interact with and more likable.
The animations also significantly increased the chatbot's perceived emotional expressivity, i.e., the ability to express human-like emotions. 
As expected, the affective animations did not have a significant impact on the chatbot's verbal communication abilities. 
Last but not the least, the animated chat balloons significantly influenced the participants' perceived personality of the chatbot in the trait of Extroversion \citep{TIPI}. These findings indicate that chat balloon animations can be effectively utilized as a design tool to shape the interaction experience and the emotional expressivity of text-based conversational agents and even configure their perceived personality traits. To date, these purposes have been mainly achieved by designing the verbal aspects of artificial agents \citep{kang2018chatbotPurpose,ruane2020personality}. Future research could thereby further explore animated chat balloons as a nonverbal approach for emotional design of conversational agents in various relevant scenarios, such as client support, or children's social-emotional learning \citep{socialEmotionalChatbot}. 

Besides answering the two research questions, our studies have shown that animated chat balloons can effectively complement existing affective communication methods in mobile or wearable social messaging, such as emojis and stickers. While these methods are often focused on conveying static facial expressions or body poses \citep{vinciarelli2009socialSignal}, animated chat balloons can capture the dynamic, temporal aspects of nonverbal interaction. \rv{Namely, the participants found that \name{} allowed them to more vividly envision the affective reactions from the other side, likely enhancing the feeling of ``face-to-face'' in the conversation, which suggests a deeper and more detailed study into how \name{} would mediate Social Presence in messaging interactions \citep{de2001socialpresence,yen2008online}.} Moreover, the participants felt that the emotional information conveyed by the chat balloon animations was intuitively received before their reading of the message content. 
This suggests that as similar to color \citep{bubbleColoring}, the animation of chat balloons might also be used as affective clues for voice messages whose content is not immediately presented to the receiver. 
Future research could explore specific contexts of voice messages, especially on smartwatches (see a subsequent work by \cite{EmoWear}), since chat balloon animations seem more compatible with small screens than certain affective cues (e.g., typefaces \citep{emotype,animatedtexts}).

\rv{For future researchers and developers to practically implement AniBalloons in their message-based systems, we envision an integration into messaging environments. In social messaging, they're incorporated directly into the keyboard interface, offering users the ability to select emotional animations while typing, with suggestions provided based on the text's semantic meaning. For customer support chatbots, AniBalloons are automatically applied in line with the bot's responses, enhancing emotional engagement without requiring user interaction for selection. In the social messaging context, a recommendation mechanism is needed to balance the richness of animation options with user convenience of selection. For instance, this mechanism could employ semantic analysis to suggest a couple of most fitting emotional animations, making it easier for users to express themselves accurately without sifting through all 30 options. This ensures a straightforward, user-friendly experience that enriches digital communication with emotional depth, allowing for both quick selection and the exploration of diverse options for those who wish to personalize their messages further.}
Last but not least, as we have found, the abstract nature of animated chat balloons allows for universal communication of emotions without being tied to specific features such as gender or skin tones (which may lead to stereotyping or exclusion of certain cultural or ethnic groups).
This interestingly surfaces a unique benefit of designing relatively ``abstract'' animations to convey emotion (as also seen in \citep{AnimoLiu}), which could be more explicitly explored. 

\rv{\textbf{Limitations.} 
Our study has faced several limitations that open avenues for future research. One such area we did not focus on is the user experience associated with browsing and selecting these animations through a keyboard interface in social messaging scenarios. This oversight suggests a promising direction for further investigation to enhance user interaction and preferences understanding.
Furthermore, the use of mock-up interfaces in our study, rather than fully functioning applications, points to the necessity of deploying these animations in real-world settings. Future work should aim to develop and implement these affective animations in longitudinal, naturalistic environments to capture more comprehensive insights into their utility and user engagement.
Additionally, our research did not address the accessibility of these animations, an aspect crucial for ensuring inclusivity. Future studies should explore how these animations can be made accessible to a wider range of users, including those with disabilities, to foster a more inclusive digital communication space.
Lastly, while our study initiates the conversation on animated chat balloons in enhancing communication, the limited diversity of conversation topics explored indicates the need for a broader examination. Future research is invited to test these animations across varied conversational scenarios, further validating their effectiveness and broadening their applicability.
In essence, our study serves as a stepping stone in the affective communication landscape, underlying a preliminary basis for more extensive explorations into animated chat balloons. By addressing these highlighted areas, future research can significantly contribute to enriching nonverbal affective communication channels for messaging-based interactions.
}

\section{Conclusion}
this paper presents an exploration of chat balloon animations as a form of affective communication support for messages. Through the design and research of AniBalloons, a set of 30 chat-balloon animations conveying six types of emotions, we aimed to understand how chat balloon animations can be designed to express emotions, and how they influence the conversation experience. %in text message-based communication. 
Our results show that 80\% of the designed animations effectively communicated the intended emotions without context from the message and covered a wide range of valence-arousal parameters. Moreover, we found that animated chat balloons could enhance perceived emotional communication quality, nonverbal information conveyance, and interpersonal closeness in mobile instant messaging. Animated chat balloons could also make a chatbot more fun to interact with and more likable, and mediate its perceived personality trait. Overall, our research suggests that animated chat balloons can be adopted as a design means to intentionally mediate particular aspects of conversation experience in text-based communications. We thereby contribute to a set of chat balloon animations that can complement nonverbal affective communication for a range of text-based interfaces and empirical insights into how animated chat balloons influence conversation experience in instant messaging and chatbot interaction.

\section*{Acknowledgements}
% We thank all our participants for their time and valuable input. 
% This work is supported in part by the Natural Sciences and Engineering Research Council of Canada (NSERC) and NSSFC Art (22CG184). 
Placeholder texts for acknowledgement after publication.


\appendix

\bibliographystyle{elsarticle-harv} 
\bibliography{references}

\begin{thebibliography}{77}
\expandafter\ifx\csname natexlab\endcsname\relax\def\natexlab#1{#1}\fi
\providecommand{\url}[1]{\texttt{#1}}
\providecommand{\href}[2]{#2}
\providecommand{\path}[1]{#1}
\providecommand{\DOIprefix}{doi:}
\providecommand{\ArXivprefix}{arXiv:}
\providecommand{\URLprefix}{URL: }
\providecommand{\Pubmedprefix}{pmid:}
\providecommand{\doi}[1]{\href{http://dx.doi.org/#1}{\path{#1}}}
\providecommand{\Pubmed}[1]{\href{pmid:#1}{\path{#1}}}
\providecommand{\bibinfo}[2]{#2}
\ifx\xfnm\relax \def\xfnm[#1]{\unskip,\space#1}\fi
%Type = Inproceedings
\bibitem[{An et~al.(2023)An, Zhang, Gao, Zhou, Du, Yan, Xiao and
  Zhao}]{Aniballoon}
\bibinfo{author}{An, P.}, \bibinfo{author}{Zhang, C.}, \bibinfo{author}{Gao,
  H.}, \bibinfo{author}{Zhou, Z.}, \bibinfo{author}{Du, L.},
  \bibinfo{author}{Yan, C.}, \bibinfo{author}{Xiao, Y.}, \bibinfo{author}{Zhao,
  J.}, \bibinfo{year}{2023}.
\newblock \bibinfo{title}{Affective affordance of message balloon animations:
  An early exploration of aniballoons}, in: \bibinfo{booktitle}{Companion
  Publication of the 2023 Conference on Computer Supported Cooperative Work and
  Social Computing}, \bibinfo{publisher}{Association for Computing Machinery},
  \bibinfo{address}{New York, NY, USA}. p. \bibinfo{pages}{138–143}.
\newblock \URLprefix \url{https://doi.org/10.1145/3584931.3607017},
  \DOIprefix\doi{10.1145/3584931.3607017}.
%Type = Inproceedings
\bibitem[{An et~al.(2022)An, Zhou, Liu, Yin, Du, Huang and Zhao}]{vibemoji}
\bibinfo{author}{An, P.}, \bibinfo{author}{Zhou, Z.}, \bibinfo{author}{Liu,
  Q.}, \bibinfo{author}{Yin, Y.}, \bibinfo{author}{Du, L.},
  \bibinfo{author}{Huang, D.Y.}, \bibinfo{author}{Zhao, J.},
  \bibinfo{year}{2022}.
\newblock \bibinfo{title}{Vibemoji: Exploring user-authoring multi-modal
  emoticons in social communication}, in: \bibinfo{booktitle}{Proceedings of
  the 2022 CHI Conference on Human Factors in Computing Systems},
  \bibinfo{publisher}{Association for Computing Machinery},
  \bibinfo{address}{New York, NY, USA}.
\newblock \URLprefix \url{https://doi.org/10.1145/3491102.3501940},
  \DOIprefix\doi{10.1145/3491102.3501940}.
%Type = Inproceedings
\bibitem[{An et~al.(2024)An, Zhu, Zhang, Yin, Ma, Yan, Du and Zhao}]{EmoWear}
\bibinfo{author}{An, P.}, \bibinfo{author}{Zhu, J.S.}, \bibinfo{author}{Zhang,
  Z.}, \bibinfo{author}{Yin, Y.}, \bibinfo{author}{Ma, Q.},
  \bibinfo{author}{Yan, C.}, \bibinfo{author}{Du, L.}, \bibinfo{author}{Zhao,
  J.}, \bibinfo{year}{2024}.
\newblock \bibinfo{title}{Emowear: Exploring emotional teasers for voice
  message interaction on smartwatches}, in: \bibinfo{booktitle}{Proceedings of
  the CHI Conference on Human Factors in Computing Systems},
  \bibinfo{publisher}{Association for Computing Machinery},
  \bibinfo{address}{New York, NY, USA}.
\newblock \URLprefix \url{https://doi.org/10.1145/3613904.3642101},
  \DOIprefix\doi{10.1145/3613904.3642101}.
%Type = Inproceedings
\bibitem[{Aoki et~al.(2022)Aoki, Chujo, Matsui, Choi and
  Hautasaari}]{emoBalloon}
\bibinfo{author}{Aoki, T.}, \bibinfo{author}{Chujo, R.},
  \bibinfo{author}{Matsui, K.}, \bibinfo{author}{Choi, S.},
  \bibinfo{author}{Hautasaari, A.}, \bibinfo{year}{2022}.
\newblock \bibinfo{title}{Emoballoon - conveying emotional arousal in text
  chats with speech balloons}, in: \bibinfo{booktitle}{Proceedings of the 2022
  CHI Conference on Human Factors in Computing Systems},
  \bibinfo{publisher}{Association for Computing Machinery},
  \bibinfo{address}{New York, NY, USA}.
\newblock \URLprefix \url{https://doi.org/10.1145/3491102.3501920},
  \DOIprefix\doi{10.1145/3491102.3501920}.
%Type = Misc
\bibitem[{Apple(2021)}]{applememoji}
\bibinfo{author}{Apple}, \bibinfo{year}{2021}.
\newblock \bibinfo{title}{Use memoji on your iphone or ipad pro}.
\newblock \URLprefix \url{https://support.apple.com/en-ca/HT208986}.
%Type = Article
\bibitem[{Aron et~al.(1992)Aron, Aron and Smollan}]{aron1992inclusion}
\bibinfo{author}{Aron, A.}, \bibinfo{author}{Aron, E.N.},
  \bibinfo{author}{Smollan, D.}, \bibinfo{year}{1992}.
\newblock \bibinfo{title}{Inclusion of other in the self scale and the
  structure of interpersonal closeness.}
\newblock \bibinfo{journal}{Journal of personality and social psychology}
  \bibinfo{volume}{63}, \bibinfo{pages}{596}.
%Type = Article
\bibitem[{Biocca et~al.(2003)Biocca, Harms and Burgoon}]{networkedMinds}
\bibinfo{author}{Biocca, F.}, \bibinfo{author}{Harms, C.},
  \bibinfo{author}{Burgoon, J.K.}, \bibinfo{year}{2003}.
\newblock \bibinfo{title}{{Toward a More Robust Theory and Measure of Social
  Presence: Review and Suggested Criteria}}.
\newblock \bibinfo{journal}{Presence: Teleoperators and Virtual Environments}
  \bibinfo{volume}{12}, \bibinfo{pages}{456--480}.
\newblock \URLprefix \url{https://doi.org/10.1162/105474603322761270},
  \DOIprefix\doi{10.1162/105474603322761270},
  \href{http://arxiv.org/abs/https://direct.mit.edu/pvar/article-pdf/12/5/456/1623957/105474603322761270.pdf}{{\tt
  arXiv:https://direct.mit.edu/pvar/article-pdf/12/5/456/1623957/105474603322761270.pdf}}.
%Type = Article
\bibitem[{Borsci et~al.(2022)Borsci, Malizia, Schmettow, Van Der~Velde,
  Tariverdiyeva, Balaji and Chamberlain}]{borsci2022CUS}
\bibinfo{author}{Borsci, S.}, \bibinfo{author}{Malizia, A.},
  \bibinfo{author}{Schmettow, M.}, \bibinfo{author}{Van Der~Velde, F.},
  \bibinfo{author}{Tariverdiyeva, G.}, \bibinfo{author}{Balaji, D.},
  \bibinfo{author}{Chamberlain, A.}, \bibinfo{year}{2022}.
\newblock \bibinfo{title}{The chatbot usability scale: The design and pilot of
  a usability scale for interaction with ai-based conversational agents}.
\newblock \bibinfo{journal}{Personal and ubiquitous computing}
  \bibinfo{volume}{26}, \bibinfo{pages}{95--119}.
%Type = Article
\bibitem[{Braun and Clarke(2006)}]{thematicAnalysis}
\bibinfo{author}{Braun, V.}, \bibinfo{author}{Clarke, V.},
  \bibinfo{year}{2006}.
\newblock \bibinfo{title}{Using thematic analysis in psychology}.
\newblock \bibinfo{journal}{Qualitative Research in Psychology}
  \bibinfo{volume}{3}, \bibinfo{pages}{77--101}.
\newblock \URLprefix
  \url{https://www.tandfonline.com/doi/abs/10.1191/1478088706qp063oa},
  \DOIprefix\doi{10.1191/1478088706qp063oa},
  \href{http://arxiv.org/abs/https://www.tandfonline.com/doi/pdf/10.1191/1478088706qp063oa}{{\tt
  arXiv:https://www.tandfonline.com/doi/pdf/10.1191/1478088706qp063oa}}.
%Type = Misc
\bibitem[{Brian(2012)}]{plato}
\bibinfo{author}{Brian}, \bibinfo{year}{2012}.
\newblock \bibinfo{title}{Plato emoticons, revisited}.
\newblock \URLprefix
  \url{http://www.platohistory.org/blog/2012/09/plato-emoticons-revisited.html}.
%Type = Misc
\bibitem[{Burge(2020)}]{newemoji2021}
\bibinfo{author}{Burge, J.}, \bibinfo{year}{2020}.
\newblock \bibinfo{title}{217 new emojis in final list for 2021}.
\newblock \URLprefix
  \url{https://blog.emojipedia.org/emoji-use-at-all-time-high/}.
%Type = Article
\bibitem[{Buschek et~al.(2018)Buschek, Hassib and Alt}]{ChatAugmentation}
\bibinfo{author}{Buschek, D.}, \bibinfo{author}{Hassib, M.},
  \bibinfo{author}{Alt, F.}, \bibinfo{year}{2018}.
\newblock \bibinfo{title}{Personal mobile messaging in context: Chat
  augmentations for expressiveness and awareness}.
\newblock \bibinfo{journal}{ACM Trans. Comput.-Hum. Interact.}
  \bibinfo{volume}{25}.
\newblock \URLprefix \url{https://doi.org/10.1145/3201404},
  \DOIprefix\doi{10.1145/3201404}.
%Type = Article
\bibitem[{Byron(2008)}]{MiscommunicationEmotion}
\bibinfo{author}{Byron, k.}, \bibinfo{year}{2008}.
\newblock \bibinfo{title}{Carrying too heavy a load? the communication and
  miscommunication of emotion by email}.
\newblock \bibinfo{journal}{Academy of Management Review} \bibinfo{volume}{33},
  \bibinfo{pages}{309--327}.
\newblock \URLprefix \url{https://doi.org/10.5465/amr.2008.31193163},
  \DOIprefix\doi{10.5465/amr.2008.31193163},
  \href{http://arxiv.org/abs/https://doi.org/10.5465/amr.2008.31193163}{{\tt
  arXiv:https://doi.org/10.5465/amr.2008.31193163}}.
%Type = Article
\bibitem[{Campbell et~al.(2021)Campbell, Blackburn, Erickson, Pelt, Anderson,
  Peters and Gililland}]{texthealthcare}
\bibinfo{author}{Campbell, K.J.}, \bibinfo{author}{Blackburn, B.E.},
  \bibinfo{author}{Erickson, J.A.}, \bibinfo{author}{Pelt, C.E.},
  \bibinfo{author}{Anderson, L.A.}, \bibinfo{author}{Peters, C.L.},
  \bibinfo{author}{Gililland, J.M.}, \bibinfo{year}{2021}.
\newblock \bibinfo{title}{Evaluating the utility of using text messages to
  communicate with patients during the covid-19 pandemic}.
\newblock \bibinfo{journal}{JAAOS Global Research \& Reviews}
  \bibinfo{volume}{5}.
%Type = Inproceedings
\bibitem[{Chen et~al.(2021a)Chen, Yan and Suk}]{bubbleColoring}
\bibinfo{author}{Chen, Q.}, \bibinfo{author}{Yan, Y.}, \bibinfo{author}{Suk,
  H.J.}, \bibinfo{year}{2021}a.
\newblock \bibinfo{title}{Bubble coloring to visualize the speech emotion}, in:
  \bibinfo{booktitle}{Extended Abstracts of the 2021 CHI Conference on Human
  Factors in Computing Systems}, \bibinfo{publisher}{Association for Computing
  Machinery}, \bibinfo{address}{New York, NY, USA}.
\newblock \URLprefix \url{https://doi.org/10.1145/3411763.3451698},
  \DOIprefix\doi{10.1145/3411763.3451698}.
%Type = Article
\bibitem[{Chen et~al.(2021b)Chen, Cao, Yao, Lu, Peng, Mei and
  Liu}]{Chen2021_emotion}
\bibinfo{author}{Chen, Z.}, \bibinfo{author}{Cao, Y.}, \bibinfo{author}{Yao,
  H.}, \bibinfo{author}{Lu, X.}, \bibinfo{author}{Peng, X.},
  \bibinfo{author}{Mei, H.}, \bibinfo{author}{Liu, X.}, \bibinfo{year}{2021}b.
\newblock \bibinfo{title}{Emoji-powered sentiment and emotion detection from
  software developers’ communication data}.
\newblock \bibinfo{journal}{ACM Trans. Softw. Eng. Methodol.}
  \bibinfo{volume}{30}.
\newblock \URLprefix \url{https://doi.org/10.1145/3424308},
  \DOIprefix\doi{10.1145/3424308}.
%Type = Inproceedings
\bibitem[{Chevalier et~al.(2016)Chevalier, Riche, Plaisant, Chalbi and
  Hurter}]{chevalier2016animations}
\bibinfo{author}{Chevalier, F.}, \bibinfo{author}{Riche, N.H.},
  \bibinfo{author}{Plaisant, C.}, \bibinfo{author}{Chalbi, A.},
  \bibinfo{author}{Hurter, C.}, \bibinfo{year}{2016}.
\newblock \bibinfo{title}{Animations 25 years later: New roles and
  opportunities}, in: \bibinfo{booktitle}{Proceedings of the international
  working conference on advanced visual interfaces}, pp.
  \bibinfo{pages}{280--287}.
%Type = Article
\bibitem[{Choi and Aizawa(2019)}]{emotype}
\bibinfo{author}{Choi, S.}, \bibinfo{author}{Aizawa, K.}, \bibinfo{year}{2019}.
\newblock \bibinfo{title}{Emotype: Expressing emotions by changing typeface in
  mobile messenger texting}.
\newblock \bibinfo{journal}{Multimedia Tools and Applications}
  \bibinfo{volume}{78}, \bibinfo{pages}{14155--14172}.
%Type = Article
\bibitem[{De~Greef and Ijsselsteijn(2001)}]{de2001socialpresence}
\bibinfo{author}{De~Greef, P.}, \bibinfo{author}{Ijsselsteijn, W.A.},
  \bibinfo{year}{2001}.
\newblock \bibinfo{title}{Social presence in a home tele-application}.
\newblock \bibinfo{journal}{CyberPsychology \& Behavior} \bibinfo{volume}{4},
  \bibinfo{pages}{307--315}.
%Type = Article
\bibitem[{Ehrhart et~al.(2009)Ehrhart, Ehrhart, Roesch, Chung-Herrera, Nadler
  and Bradshaw}]{TIPI}
\bibinfo{author}{Ehrhart, M.G.}, \bibinfo{author}{Ehrhart, K.H.},
  \bibinfo{author}{Roesch, S.C.}, \bibinfo{author}{Chung-Herrera, B.G.},
  \bibinfo{author}{Nadler, K.}, \bibinfo{author}{Bradshaw, K.},
  \bibinfo{year}{2009}.
\newblock \bibinfo{title}{Testing the latent factor structure and construct
  validity of the ten-item personality inventory}.
\newblock \bibinfo{journal}{Personality and Individual Differences}
  \bibinfo{volume}{47}, \bibinfo{pages}{900--905}.
\newblock \DOIprefix\doi{https://doi.org/10.1016/j.paid.2009.07.012}.
%Type = Article
\bibitem[{Ekman(1992a)}]{ekman1992there}
\bibinfo{author}{Ekman, P.}, \bibinfo{year}{1992}a.
\newblock \bibinfo{title}{Are there basic emotions?}
\newblock \bibinfo{journal}{Psychological Review} \bibinfo{volume}{3},
  \bibinfo{pages}{550--553}.
%Type = Article
\bibitem[{Ekman(1992b)}]{ekman1992argument}
\bibinfo{author}{Ekman, P.}, \bibinfo{year}{1992}b.
\newblock \bibinfo{title}{An argument for basic emotions}.
\newblock \bibinfo{journal}{Cognition \& emotion} \bibinfo{volume}{6},
  \bibinfo{pages}{169--200}.
%Type = Misc
\bibitem[{Facebook(2021)}]{facebookmessenger}
\bibinfo{author}{Facebook}, \bibinfo{year}{2021}.
\newblock \bibinfo{title}{Messenger}.
\newblock \URLprefix \url{https://www.messenger.com/}.
%Type = Misc
\bibitem[{Fahlman(Retrieved 2021)}]{ScottFahlman}
\bibinfo{author}{Fahlman, S.}, \bibinfo{year}{Retrieved 2021}.
\newblock \bibinfo{title}{Smiley lore :-)}.
\newblock \URLprefix \url{https://www.cs.cmu.edu/~sef/sefSmiley.htm}.
%Type = Article
\bibitem[{Foo et~al.(2021)Foo, Dunne and Holschuh}]{AffectiveGarmentFoo}
\bibinfo{author}{Foo, E.W.}, \bibinfo{author}{Dunne, L.E.},
  \bibinfo{author}{Holschuh, B.}, \bibinfo{year}{2021}.
\newblock \bibinfo{title}{User expectations and mental models for communicating
  emotions through compressive \& warm affective garment actuation}.
\newblock \bibinfo{journal}{Proc. ACM Interact. Mob. Wearable Ubiquitous
  Technol.} \bibinfo{volume}{5}.
\newblock \URLprefix \url{https://doi.org/10.1145/3448097},
  \DOIprefix\doi{10.1145/3448097}.
%Type = Article
\bibitem[{Fu et~al.(2022)Fu, Michelson, Lin, Nguyen, Tayebi and
  Hiniker}]{socialEmotionalChatbot}
\bibinfo{author}{Fu, Y.}, \bibinfo{author}{Michelson, R.},
  \bibinfo{author}{Lin, Y.}, \bibinfo{author}{Nguyen, L.K.},
  \bibinfo{author}{Tayebi, T.J.}, \bibinfo{author}{Hiniker, A.},
  \bibinfo{year}{2022}.
\newblock \bibinfo{title}{Social emotional learning with conversational agents:
  Reviewing current designs and probing parents' ideas for future ones}.
\newblock \bibinfo{journal}{Proc. ACM Interact. Mob. Wearable Ubiquitous
  Technol.} \bibinfo{volume}{6}.
\newblock \URLprefix \url{https://doi.org/10.1145/3534622},
  \DOIprefix\doi{10.1145/3534622}.
%Type = Misc
\bibitem[{Google(2021)}]{googlegboard}
\bibinfo{author}{Google}, \bibinfo{year}{2021}.
\newblock \bibinfo{title}{Gboard}.
\newblock \URLprefix
  \url{https://blog.google/products/android/emoji-kitchen-new-mashups-mixing-experience/}.
%Type = Inproceedings
\bibitem[{Griggio et~al.(2021)Griggio, Sato, Mackay and Yatani}]{Griggio2021}
\bibinfo{author}{Griggio, C.F.}, \bibinfo{author}{Sato, A.J.},
  \bibinfo{author}{Mackay, W.E.}, \bibinfo{author}{Yatani, K.},
  \bibinfo{year}{2021}.
\newblock \bibinfo{title}{Mediating intimacy with dearboard: A co-customizable
  keyboard for everyday messaging}, in: \bibinfo{booktitle}{Proceedings of the
  2021 CHI Conference on Human Factors in Computing Systems},
  \bibinfo{publisher}{Association for Computing Machinery},
  \bibinfo{address}{New York, NY, USA}.
\newblock \URLprefix \url{https://doi.org/10.1145/3411764.3445757}.
%Type = Inproceedings
\bibitem[{Hagen et~al.(2019)Hagen, Falling, Lisnichenko, Elmadany, Mehta,
  Abdul-Mageed, Costakis and Keller}]{Hagen2019_emojiuse}
\bibinfo{author}{Hagen, L.}, \bibinfo{author}{Falling, M.},
  \bibinfo{author}{Lisnichenko, O.}, \bibinfo{author}{Elmadany, A.A.},
  \bibinfo{author}{Mehta, P.}, \bibinfo{author}{Abdul-Mageed, M.},
  \bibinfo{author}{Costakis, J.}, \bibinfo{author}{Keller, T.E.},
  \bibinfo{year}{2019}.
\newblock \bibinfo{title}{Emoji use in twitter white nationalism
  communication}, in: \bibinfo{booktitle}{Conference Companion Publication of
  the 2019 on Computer Supported Cooperative Work and Social Computing},
  \bibinfo{publisher}{Association for Computing Machinery},
  \bibinfo{address}{New York, NY, USA}. p. \bibinfo{pages}{201–205}.
\newblock \URLprefix \url{https://doi.org/10.1145/3311957.3359495},
  \DOIprefix\doi{10.1145/3311957.3359495}.
%Type = Inproceedings
\bibitem[{Hancock et~al.(2007)Hancock, Landrigan and Silver}]{textEmotion}
\bibinfo{author}{Hancock, J.T.}, \bibinfo{author}{Landrigan, C.},
  \bibinfo{author}{Silver, C.}, \bibinfo{year}{2007}.
\newblock \bibinfo{title}{Expressing emotion in text-based communication}, in:
  \bibinfo{booktitle}{Proceedings of the SIGCHI Conference on Human Factors in
  Computing Systems}, \bibinfo{publisher}{Association for Computing Machinery},
  \bibinfo{address}{New York, NY, USA}. p. \bibinfo{pages}{929–932}.
\newblock \URLprefix \url{https://doi.org/10.1145/1240624.1240764},
  \DOIprefix\doi{10.1145/1240624.1240764}.
%Type = Inproceedings
\bibitem[{Harrison et~al.(2011)Harrison, Hsieh, Willis, Forlizzi and
  Hudson}]{Harrison2011}
\bibinfo{author}{Harrison, C.}, \bibinfo{author}{Hsieh, G.},
  \bibinfo{author}{Willis, K.D.}, \bibinfo{author}{Forlizzi, J.},
  \bibinfo{author}{Hudson, S.E.}, \bibinfo{year}{2011}.
\newblock \bibinfo{title}{Kineticons: Using iconographic motion in graphical
  user interface design}, in: \bibinfo{booktitle}{Proceedings of the SIGCHI
  Conference on Human Factors in Computing Systems},
  \bibinfo{publisher}{Association for Computing Machinery},
  \bibinfo{address}{New York, NY, USA}. p. \bibinfo{pages}{1999–2008}.
\newblock \URLprefix \url{https://doi.org/10.1145/1978942.1979232},
  \DOIprefix\doi{10.1145/1978942.1979232}.
%Type = Inproceedings
\bibitem[{Hautasaari et~al.(2024)Hautasaari, Aramaki, Chujo and
  Naemura}]{emoScribe}
\bibinfo{author}{Hautasaari, A.}, \bibinfo{author}{Aramaki, M.},
  \bibinfo{author}{Chujo, R.}, \bibinfo{author}{Naemura, T.},
  \bibinfo{year}{2024}.
\newblock \bibinfo{title}{Emoscribe camera: A virtual camera system to enliven
  online conferencing with automatically generated emotional text captions},
  in: \bibinfo{booktitle}{Extended Abstracts of the 2024 CHI Conference on
  Human Factors in Computing Systems}, \bibinfo{publisher}{Association for
  Computing Machinery}, \bibinfo{address}{New York, NY, USA}.
\newblock \URLprefix \url{https://doi.org/10.1145/3613905.3650987},
  \DOIprefix\doi{10.1145/3613905.3650987}.
%Type = Inproceedings
\bibitem[{Jiang et~al.(2017)Jiang, Brubaker and Fiesler}]{jiang2017_GIF}
\bibinfo{author}{Jiang, J.A.}, \bibinfo{author}{Brubaker, J.R.},
  \bibinfo{author}{Fiesler, C.}, \bibinfo{year}{2017}.
\newblock \bibinfo{title}{Understanding diverse interpretations of animated
  gifs}, in: \bibinfo{booktitle}{Proceedings of the 2017 CHI Conference
  Extended Abstracts on Human Factors in Computing Systems},
  \bibinfo{publisher}{Association for Computing Machinery},
  \bibinfo{address}{New York, NY, USA}. p. \bibinfo{pages}{1726–1732}.
\newblock \URLprefix \url{https://doi.org/10.1145/3027063.3053139},
  \DOIprefix\doi{10.1145/3027063.3053139}.
%Type = Article
\bibitem[{Kang(2018)}]{kang2018chatbotPurpose}
\bibinfo{author}{Kang, M.}, \bibinfo{year}{2018}.
\newblock \bibinfo{title}{A study of chatbot personality based on the purposes
  of chatbot}.
\newblock \bibinfo{journal}{The Journal of the Korea Contents Association}
  \bibinfo{volume}{18}, \bibinfo{pages}{319--329}.
%Type = Article
\bibitem[{Kelly and Watts(2015)}]{kelly2015characterising}
\bibinfo{author}{Kelly, R.}, \bibinfo{author}{Watts, L.}, \bibinfo{year}{2015}.
\newblock \bibinfo{title}{Characterising the inventive appropriation of emoji
  as relationally meaningful in mediated close personal relationships}.
\newblock \bibinfo{journal}{Experiences of technology appropriation:
  Unanticipated users, usage, circumstances, and design} \bibinfo{volume}{2}.
%Type = Inproceedings
\bibitem[{Kim et~al.(2020)Kim, Gong, Han, Kim, Ko and Lee}]{messageImage}
\bibinfo{author}{Kim, J.}, \bibinfo{author}{Gong, T.}, \bibinfo{author}{Han,
  K.}, \bibinfo{author}{Kim, J.}, \bibinfo{author}{Ko, J.},
  \bibinfo{author}{Lee, S.J.}, \bibinfo{year}{2020}.
\newblock \bibinfo{title}{Messaging beyond texts with real-time image
  suggestions}, in: \bibinfo{booktitle}{22nd International Conference on
  Human-Computer Interaction with Mobile Devices and Services},
  \bibinfo{publisher}{Association for Computing Machinery},
  \bibinfo{address}{New York, NY, USA}.
\newblock \URLprefix \url{https://doi.org/10.1145/3379503.3403553},
  \DOIprefix\doi{10.1145/3379503.3403553}.
%Type = Article
\bibitem[{Kluck et~al.(2021)Kluck, Stoyanova and
  Krämer}]{textBenefitInPandemic}
\bibinfo{author}{Kluck, J.P.}, \bibinfo{author}{Stoyanova, F.},
  \bibinfo{author}{Krämer, N.C.}, \bibinfo{year}{2021}.
\newblock \bibinfo{title}{Putting the social back into physical distancing: The
  role of digital connections in a pandemic crisis}.
\newblock \bibinfo{journal}{International Journal of Psychology}
  \bibinfo{volume}{56}, \bibinfo{pages}{594--606}.
\newblock \URLprefix
  \url{https://onlinelibrary.wiley.com/doi/abs/10.1002/ijop.12746},
  \DOIprefix\doi{https://doi.org/10.1002/ijop.12746},
  \href{http://arxiv.org/abs/https://onlinelibrary.wiley.com/doi/pdf/10.1002/ijop.12746}{{\tt
  arXiv:https://onlinelibrary.wiley.com/doi/pdf/10.1002/ijop.12746}}.
%Type = Inproceedings
\bibitem[{Kurlander et~al.(1996)Kurlander, Skelly and Salesin}]{comicChat}
\bibinfo{author}{Kurlander, D.}, \bibinfo{author}{Skelly, T.},
  \bibinfo{author}{Salesin, D.}, \bibinfo{year}{1996}.
\newblock \bibinfo{title}{Comic chat}, in: \bibinfo{booktitle}{Proceedings of
  the 23rd Annual Conference on Computer Graphics and Interactive Techniques},
  \bibinfo{publisher}{Association for Computing Machinery},
  \bibinfo{address}{New York, NY, USA}. p. \bibinfo{pages}{225–236}.
\newblock \URLprefix \url{https://doi.org/10.1145/237170.237260},
  \DOIprefix\doi{10.1145/237170.237260}.
%Type = Article
\bibitem[{Lan et~al.(2022)Lan, Shi, Wu, Jiao and Cao}]{kineticharts}
\bibinfo{author}{Lan, X.}, \bibinfo{author}{Shi, Y.}, \bibinfo{author}{Wu, Y.},
  \bibinfo{author}{Jiao, X.}, \bibinfo{author}{Cao, N.}, \bibinfo{year}{2022}.
\newblock \bibinfo{title}{Kineticharts: Augmenting affective expressiveness of
  charts in data stories with animation design}.
\newblock \bibinfo{journal}{IEEE Transactions on Visualization and Computer
  Graphics} \bibinfo{volume}{28}, \bibinfo{pages}{933--943}.
\newblock \DOIprefix\doi{10.1109/TVCG.2021.3114775}.
%Type = Inproceedings
\bibitem[{Lasseter(1987)}]{lasseter1987principles}
\bibinfo{author}{Lasseter, J.}, \bibinfo{year}{1987}.
\newblock \bibinfo{title}{Principles of traditional animation applied to 3d
  computer animation}, in: \bibinfo{booktitle}{Proceedings of the 14th annual
  conference on Computer graphics and interactive techniques}, pp.
  \bibinfo{pages}{35--44}.
%Type = Article
\bibitem[{Liu et~al.(2017)Liu, Dabbish and Kaufman}]{heartRateMessageLiu}
\bibinfo{author}{Liu, F.}, \bibinfo{author}{Dabbish, L.},
  \bibinfo{author}{Kaufman, G.}, \bibinfo{year}{2017}.
\newblock \bibinfo{title}{Supporting social interactions with an expressive
  heart rate sharing application}.
\newblock \bibinfo{journal}{Proc. ACM Interact. Mob. Wearable Ubiquitous
  Technol.} \bibinfo{volume}{1}.
\newblock \URLprefix \url{https://doi.org/10.1145/3130943},
  \DOIprefix\doi{10.1145/3130943}.
%Type = Article
\bibitem[{Liu et~al.(2019)Liu, Esparza, Pavlovskaia, Kaufman, Dabbish and
  Monroy-Hern\'{a}ndez}]{AnimoLiu}
\bibinfo{author}{Liu, F.}, \bibinfo{author}{Esparza, M.},
  \bibinfo{author}{Pavlovskaia, M.}, \bibinfo{author}{Kaufman, G.},
  \bibinfo{author}{Dabbish, L.}, \bibinfo{author}{Monroy-Hern\'{a}ndez, A.},
  \bibinfo{year}{2019}.
\newblock \bibinfo{title}{Animo: Sharing biosignals on a smartwatch for
  lightweight social connection}.
\newblock \bibinfo{journal}{Proc. ACM Interact. Mob. Wearable Ubiquitous
  Technol.} \bibinfo{volume}{3}.
\newblock \URLprefix \url{https://doi.org/10.1145/3314405},
  \DOIprefix\doi{10.1145/3314405}.
%Type = Inproceedings
\bibitem[{Liu et~al.(2021)Liu, Park, Tham, Tsai, Dabbish, Kaufman and
  Monroy-Hern\'{a}ndez}]{Otter}
\bibinfo{author}{Liu, F.}, \bibinfo{author}{Park, C.}, \bibinfo{author}{Tham,
  Y.J.}, \bibinfo{author}{Tsai, T.Y.}, \bibinfo{author}{Dabbish, L.},
  \bibinfo{author}{Kaufman, G.}, \bibinfo{author}{Monroy-Hern\'{a}ndez, A.},
  \bibinfo{year}{2021}.
\newblock \bibinfo{title}{Significant otter: Understanding the role of
  biosignals in communication}, in: \bibinfo{booktitle}{Proceedings of the 2021
  CHI Conference on Human Factors in Computing Systems},
  \bibinfo{publisher}{Association for Computing Machinery},
  \bibinfo{address}{New York, NY, USA}.
\newblock \URLprefix \url{https://doi.org/10.1145/3411764.3445200},
  \DOIprefix\doi{10.1145/3411764.3445200}.
%Type = Inproceedings
\bibitem[{Ma et~al.(2012)Ma, Forlizzi and Dow}]{Ma2012GuidelinesFD}
\bibinfo{author}{Ma, X.}, \bibinfo{author}{Forlizzi, J.}, \bibinfo{author}{Dow,
  S.}, \bibinfo{year}{2012}.
\newblock \bibinfo{title}{Guidelines for depicting emotions in storyboard
  scenarios}.
%Type = Article
\bibitem[{Moghaddam(2006)}]{moghaddam2006coding}
\bibinfo{author}{Moghaddam, A.}, \bibinfo{year}{2006}.
\newblock \bibinfo{title}{Coding issues in grounded theory.}
\newblock \bibinfo{journal}{Issues in educational research}
  \bibinfo{volume}{16}, \bibinfo{pages}{52--66}.
%Type = Article
\bibitem[{Nguyen et~al.(2022)Nguyen, Gruber, Marler, Hunsaker, Fuchs and
  Hargittai}]{stayConnected}
\bibinfo{author}{Nguyen, M.H.}, \bibinfo{author}{Gruber, J.},
  \bibinfo{author}{Marler, W.}, \bibinfo{author}{Hunsaker, A.},
  \bibinfo{author}{Fuchs, J.}, \bibinfo{author}{Hargittai, E.},
  \bibinfo{year}{2022}.
\newblock \bibinfo{title}{Staying connected while physically apart: Digital
  communication when face-to-face interactions are limited}.
\newblock \bibinfo{journal}{New Media \& Society} \bibinfo{volume}{24},
  \bibinfo{pages}{2046--2067}.
\newblock \URLprefix \url{https://doi.org/10.1177/1461444820985442},
  \DOIprefix\doi{10.1177/1461444820985442},
  \href{http://arxiv.org/abs/https://doi.org/10.1177/1461444820985442}{{\tt
  arXiv:https://doi.org/10.1177/1461444820985442}}.
%Type = Article
\bibitem[{Nguyen et~al.(2021)Nguyen, Hargittai and
  Marler}]{textIncreaseDuringCovid}
\bibinfo{author}{Nguyen, M.H.}, \bibinfo{author}{Hargittai, E.},
  \bibinfo{author}{Marler, W.}, \bibinfo{year}{2021}.
\newblock \bibinfo{title}{Digital inequality in communication during a time of
  physical distancing: The case of covid-19}.
\newblock \bibinfo{journal}{Computers in Human Behavior} \bibinfo{volume}{120},
  \bibinfo{pages}{106717}.
\newblock \URLprefix
  \url{https://www.sciencedirect.com/science/article/pii/S074756322100039X},
  \DOIprefix\doi{https://doi.org/10.1016/j.chb.2021.106717}.
%Type = Inproceedings
\bibitem[{Ohene-Djan et~al.(2007)Ohene-Djan, Wright and
  Combie-Smith}]{emotionalSubtitles}
\bibinfo{author}{Ohene-Djan, J.}, \bibinfo{author}{Wright, J.},
  \bibinfo{author}{Combie-Smith, K.}, \bibinfo{year}{2007}.
\newblock \bibinfo{title}{Emotional subtitles: A system and potential
  applications for deaf and hearing impaired people.}, in:
  \bibinfo{booktitle}{CVHI}.
%Type = Inproceedings
\bibitem[{Peng et~al.(2018)Peng, Hsi, Taele, Lin, Lai, Hsu, Chen, Wu, Chen,
  Tang et~al.}]{peng2018speechbubbles}
\bibinfo{author}{Peng, Y.H.}, \bibinfo{author}{Hsi, M.W.},
  \bibinfo{author}{Taele, P.}, \bibinfo{author}{Lin, T.Y.},
  \bibinfo{author}{Lai, P.E.}, \bibinfo{author}{Hsu, L.},
  \bibinfo{author}{Chen, T.c.}, \bibinfo{author}{Wu, T.Y.},
  \bibinfo{author}{Chen, Y.A.}, \bibinfo{author}{Tang, H.H.}, et~al.,
  \bibinfo{year}{2018}.
\newblock \bibinfo{title}{Speechbubbles: Enhancing captioning experiences for
  deaf and hard-of-hearing people in group conversations}, in:
  \bibinfo{booktitle}{Proceedings of the 2018 CHI Conference on Human Factors
  in Computing Systems}, pp. \bibinfo{pages}{1--10}.
%Type = Article
\bibitem[{Rimé et~al.(2020)Rimé, Bouchat, Paquot and
  Giglio}]{RIME2020127sharingEmotion}
\bibinfo{author}{Rimé, B.}, \bibinfo{author}{Bouchat, P.},
  \bibinfo{author}{Paquot, L.}, \bibinfo{author}{Giglio, L.},
  \bibinfo{year}{2020}.
\newblock \bibinfo{title}{Intrapersonal, interpersonal, and social outcomes of
  the social sharing of emotion}.
\newblock \bibinfo{journal}{Current Opinion in Psychology}
  \bibinfo{volume}{31}, \bibinfo{pages}{127--134}.
\newblock \URLprefix
  \url{https://www.sciencedirect.com/science/article/pii/S2352250X19301472},
  \DOIprefix\doi{https://doi.org/10.1016/j.copsyc.2019.08.024}.
  \bibinfo{note}{privacy and Disclosure, Online and in Social Interactions}.
%Type = Article
\bibitem[{Rodrigues et~al.(2018)Rodrigues, Prada, Gaspar, Garrido and
  Lopes}]{rodrigues2018lisbon}
\bibinfo{author}{Rodrigues, D.}, \bibinfo{author}{Prada, M.},
  \bibinfo{author}{Gaspar, R.}, \bibinfo{author}{Garrido, M.V.},
  \bibinfo{author}{Lopes, D.}, \bibinfo{year}{2018}.
\newblock \bibinfo{title}{Lisbon emoji and emoticon database (leed): Norms for
  emoji and emoticons in seven evaluative dimensions}.
\newblock \bibinfo{journal}{Behavior research methods} \bibinfo{volume}{50},
  \bibinfo{pages}{392--405}.
%Type = Inproceedings
\bibitem[{Ruane et~al.(2020)Ruane, Farrell and
  Ventresque}]{ruane2020personality}
\bibinfo{author}{Ruane, E.}, \bibinfo{author}{Farrell, S.},
  \bibinfo{author}{Ventresque, A.}, \bibinfo{year}{2020}.
\newblock \bibinfo{title}{User perception of text-based chatbot personality},
  in: \bibinfo{booktitle}{International Workshop on Chatbot Research and
  Design}, \bibinfo{organization}{Springer}. pp. \bibinfo{pages}{32--47}.
%Type = Article
\bibitem[{Russell and Barrett(1999)}]{russell1999core}
\bibinfo{author}{Russell, J.A.}, \bibinfo{author}{Barrett, L.F.},
  \bibinfo{year}{1999}.
\newblock \bibinfo{title}{Core affect, prototypical emotional episodes, and
  other things called emotion: dissecting the elephant.}
\newblock \bibinfo{journal}{Journal of personality and social psychology}
  \bibinfo{volume}{76}, \bibinfo{pages}{805}.
%Type = Article
\bibitem[{Salancik and Pfeffer(1978)}]{salancik1978social}
\bibinfo{author}{Salancik, G.R.}, \bibinfo{author}{Pfeffer, J.},
  \bibinfo{year}{1978}.
\newblock \bibinfo{title}{A social information processing approach to job
  attitudes and task design}.
\newblock \bibinfo{journal}{Administrative science quarterly} ,
  \bibinfo{pages}{224--253}.
%Type = Inproceedings
\bibitem[{Shi et~al.(2021)Shi, Lan, Li, Li and Cao}]{motionDesignSpace}
\bibinfo{author}{Shi, Y.}, \bibinfo{author}{Lan, X.}, \bibinfo{author}{Li, J.},
  \bibinfo{author}{Li, Z.}, \bibinfo{author}{Cao, N.}, \bibinfo{year}{2021}.
\newblock \bibinfo{title}{Communicating with motion: A design space for
  animated visual narratives in data videos}, in:
  \bibinfo{booktitle}{Proceedings of the 2021 CHI Conference on Human Factors
  in Computing Systems}, \bibinfo{publisher}{Association for Computing
  Machinery}, \bibinfo{address}{New York, NY, USA}.
\newblock \URLprefix \url{https://doi.org/10.1145/3411764.3445337},
  \DOIprefix\doi{10.1145/3411764.3445337}.
%Type = Inproceedings
\bibitem[{Shi et~al.(2018)Shi, Yan, Ma, Lou and Cao}]{voiceAgent}
\bibinfo{author}{Shi, Y.}, \bibinfo{author}{Yan, X.}, \bibinfo{author}{Ma, X.},
  \bibinfo{author}{Lou, Y.}, \bibinfo{author}{Cao, N.}, \bibinfo{year}{2018}.
\newblock \bibinfo{title}{Designing emotional expressions of conversational
  states for voice assistants: Modality and engagement}, in:
  \bibinfo{booktitle}{Extended Abstracts of the 2018 CHI Conference on Human
  Factors in Computing Systems}, \bibinfo{publisher}{Association for Computing
  Machinery}, \bibinfo{address}{New York, NY, USA}. p. \bibinfo{pages}{1–6}.
\newblock \URLprefix \url{https://doi.org/10.1145/3170427.3188560},
  \DOIprefix\doi{10.1145/3170427.3188560}.
%Type = Book
\bibitem[{Short et~al.(1976)Short, Williams and Christie}]{short1976social}
\bibinfo{author}{Short, J.}, \bibinfo{author}{Williams, E.},
  \bibinfo{author}{Christie, B.}, \bibinfo{year}{1976}.
\newblock \bibinfo{title}{The Social Psychology of Telecommunications}.
\newblock \bibinfo{publisher}{Wiley}.
\newblock \URLprefix \url{https://books.google.co.jp/books?id=Ze63AAAAIAAJ}.
%Type = Inbook
\bibitem[{Sonderegger et~al.(2016)Sonderegger, Heyden, Chavaillaz and
  Sauer}]{Sonderegger2016_AniSAMAniAvatar}
\bibinfo{author}{Sonderegger, A.}, \bibinfo{author}{Heyden, K.},
  \bibinfo{author}{Chavaillaz, A.}, \bibinfo{author}{Sauer, J.},
  \bibinfo{year}{2016}.
\newblock \bibinfo{title}{AniSAM \& AniAvatar: Animated Visualizations of
  Affective States}. \bibinfo{publisher}{Association for Computing Machinery},
  \bibinfo{address}{New York, NY, USA}.
\newblock p. \bibinfo{pages}{4828–4837}.
\newblock \URLprefix \url{https://doi.org/10.1145/2858036.2858365}.
%Type = Misc
\bibitem[{Telegram(2021)}]{Telegram}
\bibinfo{author}{Telegram}, \bibinfo{year}{2021}.
\newblock \bibinfo{title}{Telegram}.
\newblock \URLprefix \url{https://telegram.org/}.
%Type = Book
\bibitem[{Thomas et~al.(1995)Thomas, Johnston and Thomas}]{thomas1995illusion}
\bibinfo{author}{Thomas, F.}, \bibinfo{author}{Johnston, O.},
  \bibinfo{author}{Thomas, F.}, \bibinfo{year}{1995}.
\newblock \bibinfo{title}{The illusion of life: Disney animation}.
\newblock \bibinfo{publisher}{Hyperion New York}.
%Type = Article
\bibitem[{Thompson et~al.(2020)Thompson, Liu, Li and
  Stasko}]{animatedDataGraphics}
\bibinfo{author}{Thompson, J.}, \bibinfo{author}{Liu, Z.}, \bibinfo{author}{Li,
  W.}, \bibinfo{author}{Stasko, J.}, \bibinfo{year}{2020}.
\newblock \bibinfo{title}{Understanding the design space and authoring
  paradigms for animated data graphics}.
\newblock \bibinfo{journal}{Computer Graphics Forum} \bibinfo{volume}{39},
  \bibinfo{pages}{207--218}.
\newblock \URLprefix
  \url{https://onlinelibrary.wiley.com/doi/abs/10.1111/cgf.13974},
  \DOIprefix\doi{https://doi.org/10.1111/cgf.13974},
  \href{http://arxiv.org/abs/https://onlinelibrary.wiley.com/doi/pdf/10.1111/cgf.13974}{{\tt
  arXiv:https://onlinelibrary.wiley.com/doi/pdf/10.1111/cgf.13974}}.
%Type = Inproceedings
\bibitem[{Tieryas et~al.(2017)Tieryas, Garcia, Truman and
  Bonifacio}]{louToLife}
\bibinfo{author}{Tieryas, P.}, \bibinfo{author}{Garcia, H.},
  \bibinfo{author}{Truman, S.}, \bibinfo{author}{Bonifacio, E.},
  \bibinfo{year}{2017}.
\newblock \bibinfo{title}{Bringing lou to life: A study in creating lou}, in:
  \bibinfo{booktitle}{ACM SIGGRAPH 2017 Talks}, \bibinfo{publisher}{Association
  for Computing Machinery}, \bibinfo{address}{New York, NY, USA}.
\newblock \URLprefix \url{https://doi.org/10.1145/3084363.3085089},
  \DOIprefix\doi{10.1145/3084363.3085089}.
%Type = Inproceedings
\bibitem[{de~la Torre-Arenas and Cruz(2017)}]{de2017taxonomy}
\bibinfo{author}{de~la Torre-Arenas, I.}, \bibinfo{author}{Cruz, P.},
  \bibinfo{year}{2017}.
\newblock \bibinfo{title}{A taxonomy of motion applications in data
  visualization}, in: \bibinfo{booktitle}{Proceedings of the symposium on
  Computational Aesthetics}, pp. \bibinfo{pages}{1--2}.
%Type = Article
\bibitem[{Vinciarelli et~al.(2009)Vinciarelli, Pantic and
  Bourlard}]{vinciarelli2009socialSignal}
\bibinfo{author}{Vinciarelli, A.}, \bibinfo{author}{Pantic, M.},
  \bibinfo{author}{Bourlard, H.}, \bibinfo{year}{2009}.
\newblock \bibinfo{title}{Social signal processing: Survey of an emerging
  domain}.
\newblock \bibinfo{journal}{Image and vision computing} \bibinfo{volume}{27},
  \bibinfo{pages}{1743--1759}.
%Type = Article
\bibitem[{Walther(1992)}]{waltherSIP92}
\bibinfo{author}{Walther, J.B.}, \bibinfo{year}{1992}.
\newblock \bibinfo{title}{Interpersonal effects in computer-mediated
  interaction: A relational perspective}.
\newblock \bibinfo{journal}{Communication Research} \bibinfo{volume}{19},
  \bibinfo{pages}{52--90}.
\newblock \URLprefix \url{https://doi.org/10.1177/009365092019001003},
  \DOIprefix\doi{10.1177/009365092019001003},
  \href{http://arxiv.org/abs/https://doi.org/10.1177/009365092019001003}{{\tt
  arXiv:https://doi.org/10.1177/009365092019001003}}.
%Type = Inbook
\bibitem[{Walther(2015)}]{SIPTheory}
\bibinfo{author}{Walther, J.B.}, \bibinfo{year}{2015}.
\newblock \bibinfo{title}{Social Information Processing Theory (CMC)}.
  \bibinfo{publisher}{John Wiley \& Sons, Ltd}.
\newblock pp. \bibinfo{pages}{1--13}.
\newblock \URLprefix
  \url{https://onlinelibrary.wiley.com/doi/abs/10.1002/9781118540190.wbeic192},
  \DOIprefix\doi{https://doi.org/10.1002/9781118540190.wbeic192},
  \href{http://arxiv.org/abs/https://onlinelibrary.wiley.com/doi/pdf/10.1002/9781118540190.wbeic192}{{\tt
  arXiv:https://onlinelibrary.wiley.com/doi/pdf/10.1002/9781118540190.wbeic192}}.
%Type = Article
\bibitem[{Walther and D’Addario(2001)}]{emoticonCommunication}
\bibinfo{author}{Walther, J.B.}, \bibinfo{author}{D’Addario, K.P.},
  \bibinfo{year}{2001}.
\newblock \bibinfo{title}{The impacts of emoticons on message interpretation in
  computer-mediated communication}.
\newblock \bibinfo{journal}{Social Science Computer Review}
  \bibinfo{volume}{19}, \bibinfo{pages}{324--347}.
\newblock \URLprefix \url{https://doi.org/10.1177/089443930101900307},
  \DOIprefix\doi{10.1177/089443930101900307},
  \href{http://arxiv.org/abs/https://doi.org/10.1177/089443930101900307}{{\tt
  arXiv:https://doi.org/10.1177/089443930101900307}}.
%Type = Inproceedings
\bibitem[{Wang et~al.(2004a)Wang, Prendinger and Igarashi}]{animatedtexts}
\bibinfo{author}{Wang, H.}, \bibinfo{author}{Prendinger, H.},
  \bibinfo{author}{Igarashi, T.}, \bibinfo{year}{2004}a.
\newblock \bibinfo{title}{Communicating emotions in online chat using
  physiological sensors and animated text}, in: \bibinfo{booktitle}{CHI '04
  Extended Abstracts on Human Factors in Computing Systems},
  \bibinfo{publisher}{Association for Computing Machinery},
  \bibinfo{address}{New York, NY, USA}. p. \bibinfo{pages}{1171–1174}.
\newblock \URLprefix \url{https://doi.org/10.1145/985921.986016},
  \DOIprefix\doi{10.1145/985921.986016}.
%Type = Inproceedings
\bibitem[{Wang et~al.(2004b)Wang, Prendinger and
  Igarashi}]{EmotionChatPhysiological}
\bibinfo{author}{Wang, H.}, \bibinfo{author}{Prendinger, H.},
  \bibinfo{author}{Igarashi, T.}, \bibinfo{year}{2004}b.
\newblock \bibinfo{title}{Communicating emotions in online chat using
  physiological sensors and animated text}, in: \bibinfo{booktitle}{CHI '04
  Extended Abstracts on Human Factors in Computing Systems},
  \bibinfo{publisher}{Association for Computing Machinery},
  \bibinfo{address}{New York, NY, USA}. p. \bibinfo{pages}{1171–1174}.
\newblock \URLprefix \url{https://doi.org/10.1145/985921.986016},
  \DOIprefix\doi{10.1145/985921.986016}.
%Type = Inbook
\bibitem[{Wiseman and Gould(2018)}]{Wiseman2018_repurposingemoji}
\bibinfo{author}{Wiseman, S.}, \bibinfo{author}{Gould, S.J.J.},
  \bibinfo{year}{2018}.
\newblock \bibinfo{title}{Repurposing Emoji for Personalised Communication: Why
  pizza-emoji Means ``I Love You''}. \bibinfo{publisher}{Association for
  Computing Machinery}, \bibinfo{address}{New York, NY, USA}.
\newblock p. \bibinfo{pages}{1–10}.
\newblock \URLprefix \url{https://doi.org/10.1145/3173574.3173726}.
%Type = Inproceedings
\bibitem[{Xie et~al.(2023)Xie, Zhou, Yu, Wang, Qu and Chen}]{WakeyWakey}
\bibinfo{author}{Xie, L.}, \bibinfo{author}{Zhou, Z.}, \bibinfo{author}{Yu,
  K.}, \bibinfo{author}{Wang, Y.}, \bibinfo{author}{Qu, H.},
  \bibinfo{author}{Chen, S.}, \bibinfo{year}{2023}.
\newblock \bibinfo{title}{Wakey-wakey: Animate text by mimicking characters in
  a gif}, in: \bibinfo{booktitle}{Proceedings of the 36th Annual ACM Symposium
  on User Interface Software and Technology}, \bibinfo{publisher}{Association
  for Computing Machinery}, \bibinfo{address}{New York, NY, USA}.
\newblock \URLprefix \url{https://doi.org/10.1145/3586183.3606813},
  \DOIprefix\doi{10.1145/3586183.3606813}.
%Type = Article
\bibitem[{Yamanishi et~al.(2017)Yamanishi, Tanaka, Nishihara and
  Fukumoto}]{balloonShape}
\bibinfo{author}{Yamanishi, R.}, \bibinfo{author}{Tanaka, H.},
  \bibinfo{author}{Nishihara, Y.}, \bibinfo{author}{Fukumoto, J.},
  \bibinfo{year}{2017}.
\newblock \bibinfo{title}{Speech-balloon shapes estimation for emotional text
  communication}.
\newblock \bibinfo{journal}{Information Engineering Express}
  \bibinfo{volume}{3}, \bibinfo{pages}{1--10}.
%Type = Article
\bibitem[{Yen and Tu(2008)}]{yen2008online}
\bibinfo{author}{Yen, C.J.}, \bibinfo{author}{Tu, C.H.}, \bibinfo{year}{2008}.
\newblock \bibinfo{title}{Online social presence: A study of score validity of
  the computer-mediated communication questionnaire}.
\newblock \bibinfo{journal}{Quarterly Review of Distance Education}
  \bibinfo{volume}{9}.
%Type = Inproceedings
\bibitem[{Zhang et~al.(2017)Zhang, Igo, Facciotti and
  Karger}]{Zhang2017_affectivestates}
\bibinfo{author}{Zhang, A.X.}, \bibinfo{author}{Igo, M.},
  \bibinfo{author}{Facciotti, M.}, \bibinfo{author}{Karger, D.},
  \bibinfo{year}{2017}.
\newblock \bibinfo{title}{Using student annotated hashtags and emojis to
  collect nuanced affective states}, in: \bibinfo{booktitle}{Proceedings of the
  Fourth (2017) ACM Conference on Learning @ Scale},
  \bibinfo{publisher}{Association for Computing Machinery},
  \bibinfo{address}{New York, NY, USA}. p. \bibinfo{pages}{319–322}.
\newblock \URLprefix \url{https://doi.org/10.1145/3051457.3054014},
  \DOIprefix\doi{10.1145/3051457.3054014}.
%Type = Article
\bibitem[{Zhang et~al.(2024)Zhang, Yu, Hu, Li and An}]{notUpset}
\bibinfo{author}{Zhang, N.}, \bibinfo{author}{Yu, B.}, \bibinfo{author}{Hu,
  J.}, \bibinfo{author}{Li, M.}, \bibinfo{author}{An, P.},
  \bibinfo{year}{2024}.
\newblock \bibinfo{title}{I'm not upset–i get it: Effects of co-workers'
  stress cues on help-seekers' social diction and empathy in telecommuting}.
\newblock \bibinfo{journal}{International Journal of Human-Computer Studies}
  \bibinfo{volume}{185}, \bibinfo{pages}{103218}.
\newblock \URLprefix
  \url{https://www.sciencedirect.com/science/article/pii/S1071581924000028},
  \DOIprefix\doi{https://doi.org/10.1016/j.ijhcs.2024.103218}.
%Type = Inbook
\bibitem[{Zhou et~al.(2017)Zhou, Hentschel and Kumar}]{zhou2017_wechat}
\bibinfo{author}{Zhou, R.}, \bibinfo{author}{Hentschel, J.},
  \bibinfo{author}{Kumar, N.}, \bibinfo{year}{2017}.
\newblock \bibinfo{title}{Goodbye Text, Hello Emoji: Mobile Communication on
  WeChat in China}. \bibinfo{publisher}{Association for Computing Machinery},
  \bibinfo{address}{New York, NY, USA}.
\newblock p. \bibinfo{pages}{748–759}.
\newblock \URLprefix \url{https://doi.org/10.1145/3025453.3025800}.
%Type = Inproceedings
\bibitem[{Zimmerman et~al.(2007)Zimmerman, Forlizzi and
  Evenson}]{Zimmerman_RtD}
\bibinfo{author}{Zimmerman, J.}, \bibinfo{author}{Forlizzi, J.},
  \bibinfo{author}{Evenson, S.}, \bibinfo{year}{2007}.
\newblock \bibinfo{title}{Research through design as a method for interaction
  design research in hci}, in: \bibinfo{booktitle}{Proceedings of the SIGCHI
  Conference on Human Factors in Computing Systems},
  \bibinfo{publisher}{Association for Computing Machinery},
  \bibinfo{address}{New York, NY, USA}. p. \bibinfo{pages}{493–502}.
\newblock \URLprefix \url{https://doi.org/10.1145/1240624.1240704},
  \DOIprefix\doi{10.1145/1240624.1240704}.

\end{thebibliography}





% \end{thebibliography}

\end{document}